\documentclass[journal]{IEEEtran}

\usepackage[numbers,sort&compress]{natbib}
\usepackage{url}
\usepackage{bm}
\usepackage{amsfonts}
\usepackage{amsmath}
\usepackage{amssymb}
\usepackage{color}

\usepackage{array,color}
\usepackage{tabularx}
\usepackage{multirow}
\usepackage{booktabs}
\usepackage{amsmath}  
\usepackage{makecell}  

\usepackage{graphicx}  
\usepackage{subfig}
\usepackage{epstopdf}
\usepackage{graphics}
\usepackage{epsfig}
\usepackage{psfrag}
\usepackage{overpic}

 \usepackage{stfloats}
 \usepackage{float}   

 \usepackage{indentfirst} 

\begin{document}

\newcommand {\Data} [1]{\mbox{${#1}$}}  

\newcommand {\DataN} [2]{\Data{\Power{{#1}}{{|{#2}}}}}  
\newcommand {\DataIJ} [3]{\Data{\Power{#1}{{|{#2}\!\times{}\!{#3}}}}}  

\newcommand {\DatassI} [2]{\!\Data{\Index{#1}{\!\Data 1},\!\Index{#1}{\!\Data 2},\!\cdots,\!\Index{#1}{\!{#2}}}}  
\newcommand {\DatasI} [2]{\Data{\Index{#1}{\Data 1},\Index{#1}{\Data 2},\cdots,\Index{#1}{#2},\cdots}}   
\newcommand {\DatasII} [3]{\Data{\Index{#1}{{\Index{#2}{\Data 1}}},\Index{#1}{{\Index{#2}{\Data 2}}},\cdots,\Index{#1}{{\Index{#2}{#3}}},\cdots}}  

\newcommand {\DatasNTt}[3]{\Data{\Index{#1}{{#2}{\Data 1}},\Index{#1}{{#2}{\Data 2}},\cdots,\Index{#1}{{#2}{#3}}} } 
\newcommand {\DatasNTn}[3]{\Data{\Index{#1}{{\Data 1}{#3}},\Index{#1}{{\Data 2}{#3}},\cdots,\Index{#1}{{#2}{#3}}} } 

\newcommand {\Vector} [1]{\Data {\mathbf {#1}}}
\newcommand {\Rdata} [1]{\Data {\hat {#1}}}
\newcommand {\Tdata} [1]{\Data {\tilde {#1}}} 
\newcommand {\Udata} [1]{\Data {\overline {#1}}} 
\newcommand {\Fdata} [1]{\Data {\mathbb {#1}}} 
\newcommand {\Prod} [2]{\Data {\prod_{\SI {#1}}^{\SI {#2}}}}  
\newcommand {\Sum} [2]{\Data {\sum_{\SI {#1}}^{\SI {#2}}}}   
\newcommand {\Belong} [2]{\Data{ {#1} \in{}{#2}}}  

\newcommand {\Abs} [1]{\Data{ \lvert {#1} \rvert}}  
\newcommand {\Mul} [2]{\Data{ {#1} \times {#2}}}  
\newcommand {\Muls} [2]{\Data{ {#1} \! \times \!{#2}}}  
\newcommand {\Div} [2]{\Data{ \frac{#1}{#2}}}  
\newcommand {\Trend} [2]{\Data{ {#1}\rightarrow{#2}}}  
\newcommand {\Sqrt} [1]{\Data {\sqrt {#1}}} 
\newcommand {\Sqrtn} [2]{\Data {\sqrt[2]{#1}}} 

\newcommand {\Power} [2]{\Data{ {#1}^{\TI {#2}}}}  
\newcommand {\Index} [2]{\Data{ {#1}_{\TI {#2}}}}  

\newcommand {\Equ} [2]{\Data{ {#1} = {#2}}}  
\newcommand {\Equs} [2]{\Data{ {#1}\! =\! {#2}}}  
\newcommand {\Equss} [3]{\Equs {#1}{\Equs {#2}{#3}}}  

\newcommand {\Equu} [2]{\Data{ {#1} \equiv {#2}}}  

\newcommand {\LE}[0] {\leqslant}
\newcommand {\GE}[0] {\geqslant}
\newcommand {\INF}[0] {\infty}
\newcommand {\MIN}[0] {\min}
\newcommand {\MAX}[0] {\max}

\newcommand {\Funcfx} [2]{\Data{ {#1}({#2})}}  
\newcommand {\Funcfzx} [3]{\Data{ {\Index {#1}{#2}}({#3})}}  
\newcommand {\Funcfnzx} [4]{\Data{ {\Index {\Power{#1}{#2}}{#3}}({#4})}}  
\newcommand {\SI}[1] {\small{#1}}
\newcommand {\TI}[1] {\tiny {#1}}
\newcommand {\Text}[1] {\text {#1}}

\newcommand {\VtS}[0]{\Index {t}{\Text {s}}}
\newcommand {\Vti}[0]{\Index {t}{i}}
\newcommand {\Vt}[0]{\Data {t}}
\newcommand {\VMSR}[0]{\Index {\kappa} {\SI{\Index {}{ \Text{MSR}}}}}
\newcommand {\VMSRmin}[0]{\Index {\kappa} {\SI{\Index {}{ \Text{min\_MSR}}}}}
\newcommand {\VAT}[0]{\Index {\Vector A}{\Text{Time}}}
\newcommand {\VPbus}[1]{\Index {P}{\Text{Bus-}{#1}}}
\newcommand {\VPbusmax}[1]{\Index {P}{\Text{max\_Bus-}{#1}}}

\newcommand {\EtS}[2]{\Equs {\Index {t}{\Text {s}}}{#1} {#2}}
\newcommand {\Eti}[2]{\Equs {\Index {t}{i}}{#1} {#2}}
\newcommand {\Et}[2]{\Equs {t}{#1} {#2}}
\newcommand {\EMSR}[2]{\Equs {\Index {\kappa} {\SI{\Index {}{ \Text{MSR}}}}}{#1} {#2}}
\newcommand {\EMSRmin}[2]{\Equs {\Index {\kappa} {\SI{\Index {}{ \Text{MSR}}}}}{#1} {#2}}

\newcommand {\EAT}[2]{\Equs {\Index {\Vector A}{\Text{Time}}}{#1} {#2}}
\newcommand {\EPbus}[3]{\Equs {\Index {P}{\Text{Bus-}{#1}}}{#2} \Text{ #3}}
\newcommand {\EPbusmax}[3]{\Equs {\Index {P}{\Text{max\_Bus-}{#1}}}{#2} \Text{ #3}}

\newcommand {\Vgam}[1]{\Index {\gamma}{#1}}
\newcommand {\Egam}[2]{\Equs {\Vgam{#1}}{#2}}

\newcommand {\Emu}[2]{\Equs {{\mu}{#1}}{#2}}
\newcommand {\Esigg}[2]{\Equs {{\sigma}^2{#1}}{#2}}

\newcommand {\Vlambda}[1]{\Index {\lambda}{#1}}

\newcommand {\VV}[1]{\Index {\Vector V}{#1}}
\newcommand {\VX}[1]{\Index {\Vector X}{#1}}
\newcommand {\Vx}[1]{\Index {\Vector x}{\SI{\Index {}{#1}}}}
\newcommand {\Vsx}[1]{\Index {x}{\SI{\Index {}{#1}}}}
\newcommand {\VZ}[1]{\Index {\Vector Z}{#1}}
\newcommand {\Vz}[1]{\Index {\Vector z}{\SI{\Index {}{#1}}}}
\newcommand {\Vsz}[1]{\Index {z}{\SI{\Index {}{#1}}}}
\newcommand {\VIndex}[2]{\Index {\Vector {#1}}{#2}}

\newcommand {\VRV}[1]{\Index {\Rdata {\Vector V}}{#1}}
\newcommand {\VRsV}[1]{\Index {\Rdata {V}}{#1}}
\newcommand {\VRX}[1]{\Index {\Rdata {\Vector X}}{#1}}
\newcommand {\VRx}[1]{\Index {\Rdata {\Vector x}}{\SI{\Index {}{#1}}}}
\newcommand {\VRsx}[1]{\Index {\Rdata {x}}{\SI{\Index {}{#1}}}}
\newcommand {\VRZ}[1]{\Index {\Rdata {\Vector Z}}{#1}}
\newcommand {\VRz}[1]{\Index {\Rdata {\Vector z}}{\SI{\Index {}{#1}}}}
\newcommand {\VRsz}[1]{\Index {\Rdata {z}}{\SI{\Index {}{#1}}}}
\newcommand {\VTX}[1]{\Index {\Tdata {\Vector X}}{#1}}
\newcommand {\VTx}[1]{\Index {\Tdata {\Vector x}}{\SI{\Index {}{#1}}}}
\newcommand {\VTsx}[1]{\Index {\Tdata {x}}{\SI{\Index {}{#1}}}}
\newcommand {\VTZ}[1]{\Index {\Tdata {\Vector Z}}{#1}}
\newcommand {\VTz}[1]{\Index {\Tdata {\Vector z}}{\SI{\Index {}{#1}}}}
\newcommand {\VTsz}[1]{\Index {\Tdata {z}}{\SI{\Index {}{#1}}}}
\newcommand {\VOG}[1]{\Vector{\Omega}{#1}}

\newcommand {\VY}[1]{\Index {\Vector Y}{#1}}
\newcommand {\Vy}[1]{\Index {\Vector y}{#1}}
\newcommand {\Vsy}[1]{\Index {y}{\SI{\Index {}{#1}}}}

\newcommand {\Sigg}[1]{\Data {\sigma}^2({#1})}
\newcommand {\Sig}[1]{\Data {\sigma}({#1})}

\newcommand {\Mu}[1]{\Data {\mu} ({#1})}
\newcommand {\Eig}[1]{\Data {\lambda}({\Vector {#1}}) }
\newcommand {\Her}[1]{\Power {#1}{\!H}}
\newcommand {\Tra}[1]{\Power {#1}{\!T}}

\newcommand {\VF}[3] {\DataIJ {\Fdata {#1}}{#2}{#3}}
\newcommand {\VRr}[2] {\DataN {\Fdata {#1}}{#2}}

\newcommand {\Tcol}[2] {\multicolumn{1}{#1}{#2} }
\newcommand {\Tcols}[3] {\multicolumn{#1}{#2}{#3} }
\newcommand {\Cur}[2] {\mbox {\Data {#1}--\Data {#2}}}

\newcommand {\VDelta}[1] {\Data {\Delta\!{#1}}}

\def \FuncC #1#2{
\begin{equation}
{#2}
\label {#1}
\end{equation}
}

\def \FuncCC #1#2#3#4#5#6{
\begin{equation}
#2=
\begin{cases}
    #3 & #4 \\
    #5 & #6
\end{cases}
\label{#1}
\end{equation}
}

\def \Figff #1#2#3#4#5#6#7{   
\begin{figure}[#7]
\centering
\subfloat[#2]{
\label{#1a}
\includegraphics[width=0.23\textwidth]{#4}
}
\subfloat[#3]{
\label{#1b}
\includegraphics[width=0.23\textwidth]{#5}
}
\caption{\small #6}
\label{#1}
\end{figure}
}

\def \Figffb #1#2#3#4#5#6#7#8#9{   
\begin{figure}[#9]
\centering
\subfloat[#2]{
\label{#1a}
\includegraphics[width=0.23\textwidth]{#5}
}
\subfloat[#3]{
\label{#1b}
\includegraphics[width=0.23\textwidth]{#6}
}

\subfloat[{#4}]{
\label{#1c}
\includegraphics[width=0.48\textwidth]{#7}
}
\caption{\small #8}
\label{#1}
\end{figure}
}

\def \Figffp #1#2#3#4#5#6#7{   
\begin{figure*}[#7]
\centering
\subfloat[#2]{
\label{#1a}
\begin{minipage}[t]{0.24\textwidth}
\centering
\includegraphics[width=1\textwidth]{#4}
\end{minipage}
}
\subfloat[#3]{
\label{#1b}
\begin{minipage}[t]{0.24\textwidth}
\centering
\includegraphics[width=1\textwidth]{#5}
\end{minipage}
}
\caption{\small #6}
\label{#1}
\end{figure*}
}

\def \Figf #1#2#3#4{   
\begin{figure}[#4]
\centering
\includegraphics[width=0.48\textwidth]{#2}

\caption{\small #3}
\label{#1}
\end{figure}
}

\def \Figfp #1#2#3#4{   
\begin{figure*}[#4]
\centering
\includegraphics[width=0.95\textwidth]{#2}

\caption{\small #3}
\label{#1}
\end{figure*}
}

\definecolor{Orange}{RGB}{249,106,027}
\definecolor{sOrange}{RGB}{251,166,118}
\definecolor{ssOrange}{RGB}{254,213,190}

\definecolor{Blue}{RGB}{008,161,217}
\definecolor{sBlue}{RGB}{090,206,249}
\definecolor{ssBlue}{RGB}{200,239,253}

\title{A Big Data Architecture Design for Smart Grids Based on Random Matrix Theory}
%
%
%

\author{Xing~He, Qian~Ai, ~\IEEEmembership{Member,~IEEE}, Robert~C. Qiu,~\IEEEmembership{Fellow,~IEEE},
Wentao~Huang, Longjian~Piao, Haichun~Liu
\thanks{This work was supported by National Key Technology Research and Development Program of Science and Technology (2013BAA01B04).}
\thanks{Xing~He, Qian~Ai, Wentao~Huang, Longjian~Piao, Haichun~Liu are with the Department of Electrical Engineering, Research Center for Big Data Engineering Technology, State Energy Smart Grid R $\&$ D Center, Shanghai Jiaotong University, Shanghai 200240, China (e-mail: {hexing\_hx@126.com)}}
\thanks{Robert~C.~Qiu is also with the Department of Electrical and Computer Engineering,
Tennessee Technological University, Cookeville, TN 38505 USA (e-mail: {rqiu@tntech.edu})}}

\maketitle


\begin{abstract}
 Model-based analysis tools, built on assumptions and simplifications, are difficult to handle smart grids with data characterized by volume, velocity, variety, and veracity (i.e. 4Vs data). This paper, using random matrix theory (RMT), motivates data-driven tools to perceive the complex grids in high-dimension; meanwhile, an architecture with detailed procedures is proposed.
 In algorithm perspective, the architecture performs a high-dimensional analysis, and compares the findings with RMT predictions to conduct anomaly detections. Mean Spectral Radius (MSR), as a statistical indicator, is defined to reflect the correlations of system data in different dimensions. In management mode perspective, a group-work mode is discussed for smart grids operation. This mode breaks through regional limitations for energy flows and data flows, and makes advanced big data analyses possible.
 For a specific large-scale zone-dividing system  with multiple connected utilities, each site, operating under the group-work mode, is able to work out the regional MSR only with its own measured/simulated data. The large-scale interconnected system, in this way, is naturally decoupled from statistical parameters perspective, rather than from engineering models perspective. Furthermore, a comparative analysis of these distributed MSRs, even with imperceptible different raw data, will produce a contour line to detect the event and locate the source. It demonstrates that the architecture is compatible with the block calculation only using the regional small database; beyond that, this architecture, as a data-driven solution, is sensitive to system situation awareness, and practical for real large-scale interconnected systems.
 Five case studies and their visualizations validate the designed architecture in various fields of power systems. To our best knowledge, this study is the first attempt to apply big data technology into smart grids.
\end{abstract}

\begin{IEEEkeywords}
big~data, smart~grid, architecture, random~matrix, high-dimension, mean~spectral~radius, large-scale~distributed~systems, group-work~mode
\end{IEEEkeywords}

%
\IEEEpeerreviewmaketitle

\section{Introduction}
\IEEEPARstart{B}{ig} data technology is a new scientific trend \cite{nature2008bigd,science2011bigd}. Driven by data analysis in high-dimension, big data technology works out data correlations (indicated by statistical parameters) to gain insight to the inherent mechanisms.
Data-driven results only rely on an unrestrained selection of system raw data (the space can be whole system or only a region, the time can be long or short, and the size can be large or small) and a general statistical procedure (for data processing).
On the other side, procedures for traditional model-based analysis, especially decoupling a practical interconnected system, are always based on assumptions and simplifications. Model-based results rely on identified causalities, specific parameters, sample selections, and training processes; imprecise or incomplete formulas/expressions, biased sample selections, and improper training processes will all lead to bad results. The results are often barely satisfied or even unsatisfied as the system size grows and complexity increases.
Generally speaking, data-driven analysis tools, rather than model-based ones, are more suitable to complex large-scale interconnected systems with readily accessible data.

Data in power systems have increased dramatically, leaving gaps and challenges; data processing is a major concern and its urgency increases with data growth. The 4Vs data (data with features of volume, variety, velocity, and veracity) \cite{IBM2014fourv} in smart grids, which can hardly be handled within a “tolerable elapsed time or hardware resources” by traditional model-based tools, have encouraged the development of an emerging paradigm---big data technology for power systems \mbox{\cite{qiu2015smart,IBM2009Manag,kezunovic2013role}}. Big data technology does not conflict with classical analyses or pretreatments. Actually, big data technology has already been successfully applied as a powerful data-driven tool for numerous phenomena, such as quantum systems \cite{brody1981random}, financial systems \cite{laloux2000random,chen2012business}, biological systems \cite{howe2008big}, as well as wireless communication networks \mbox{\cite{qiu2013bookcogsen, qiu2014Intial70N, qiu2014MIMO}}. Major tasks of the architecture for these applications seem similar---1) big data modeling, and 2) big data analysis. It is believed that big data technology will also have a wide applied scope in power systems, and the results will be fruitful.

\subsection {Contribution}
This paper, aiming to apply big data technology into smart grids, proposes a feasible architecture with detailed procedures. Firstly, we introduce random matrix theory (RMT) as our mathematical foundations. According to RMT, a standard random matrix is systematically formed to map the system. Then, we conduct the high-dimensional analysis and compare the findings (i.e. Empirical Spectrum Density, and Kernel Density Estimation) with the RMT theoretical predictions (i.e. Marchenko-Pastur Law, and Ring Law) to tell signals from white noises. Within the above mathematical procedure, a high-dimensional statistic---mean spectral radius (MSR)---is proposed to indicate the data correlations. More than that, MSR also clarifies the parameter interchanged among the utilities under the group-work mode for distributed calculation. In addition, power grids in three different periods, involving their features, management modes, and data flows and energy flows, are summarized as general backgrounds and foundations for applying big data in power systems.

Based on the architecture, we also conduct five case studies and summarize the most interesting results. 1) The comparisons between the experimental findings and the RMT predictions, as well as the proposed indicator MSR, are sensitive to event detections. In addition, data in different dimensions are correlative under high-dimensional perspective. 2) The MSR is somehow qualitatively correlated with the quantitative parameters of system performance. 3) The architecture, besides for event detections, can also be used as a new method to find the critical active power point at any bus node, taking account of probable grid fluctuations. 4) The architecture is compatible with the block calculation only using the regional small database, and practical for real large-scale distributed systems. In addition, the high-dimensional comparative analysis is sensitive to situation awareness for grids operation, even with imperceptible different measured data. 5) The architecture is suitable not only for the power flow analysis, but also for the fault detection. To our best knowledge, our study represents the first such attempt in the literature on power systems.

\subsection {Related Work}
It is well-established that data, as a vital resource, should be utilized much more efficiently in power systems. References  \cite {phadke2008wide,terzija2011wide,xie2012distributed} showed the improvement in wide-area monitoring, protection and control (WAMPAC) with utilizing PMUs data. Kanao et al. proposed a practical data utilization method based on harmonic state-estimation (HSE) for power system harmonic analysis \cite{kanao2005power}. It was a data processing method in a specific field and only available when the engineering model is accurate. Alahakoon et al. proposed advanced analytic refer to a number of techniques in many specific  fields: data mining tools, knowledge discovery tools, machine learning technologies, and so on \cite{alahakoon2013advanced}. Recently, Xu initiated power disturbance data analytics to explore useful aspects of power quality monitoring data ,and showed a wide applied scope in the future \cite{xu2013power}. The mathematical foundations and system frameworks were missing yet. For data utilization methods in power systems, although many researches, especially those methods based on specific physical models, were done in various fields, little attention has been paid to the design of a universal architecture, which is based on solid mathematical foundations and statistical procedures.

\section{Big Data, Random Matrix Theory, and Data Processing Procedure}
The nomenclature is given as Table \ref{tab:Nomenclature}.

\begin{table}[htbp]
\caption{Some Frequently Used Notations in the Theory}
\label{tab:Nomenclature}
\centering

\begin{tabularx}{0.48\textwidth} { l !{\color{black}\vrule width1pt} p{6.8cm} } 

\toprule[1.5pt]
\hline
\textit{Notations} & \textit{Means}\\

\Xhline{1pt}

\VX{},\Vx{},\Vsx,\Vsx{i\!,j} & a matrix, a vector, a single value, an entry of a matrix \\

\VRX{},\VRx{},\VRsx{} & hat: raw data\\
\VTX{},\VTx{},\VTsx{},\VTZ{} & tilde: transformation data, formed by normalization\\
\Udata{\VRx{i}},\Udata{\VTx{i}} & overline: average\\

\Data{N,T,c} & the numbers of rows and columns, \Equs{c}{N/T}\\
\VF {C}{N}{T} & \Muls{N}{T} dimensional complex space\\
\VX{u} &  the singular value equivalent of the matrix \VTX{}\\

\Vector S & Covariance matrix of \VX{}: \Belong {\Equs {\Vector S}{\Div 1 N\VX{}\Her{\VX{}}}}{\VF CNN}\\

\VZ{},\Data{L} & \Data{L} independent matrices product: \Data {\VZ{}=\Prod{i=1}{L} \VX {u,i}}\\

\Vlambda{\Vector S}, \Vlambda{\Vector Z} & the eigenvalue of matrix \Vector S, \VZ{}\\

\Vlambda{\Vector S,i} & the \Data i-th eigenvalue of matrix \Vector S\\
\Data{r} & the circle radius on the complex plane of eigenvalues\\
\VMSR{} &  mean value of radius for all the eigenvalues of \VTZ{}: {\Udata {\Vector r_{\Vlambda{\VTZ{}}}}}\\

\Mu{\Vsx{}},\Data{\Sigg{\Vsx{}}} & mean, variance for \Vsx{}\\

\toprule[1pt]

\end{tabularx}

\end{table}
Big data is an emerging paradigm applied to datasets whose size is beyond the ability of commonly used software tools to capture, manage, and process the data within a tolerable elapsed time.  Various technologies are being discussed to support the handling of big data such as massively parallel processing databases, scalable storage systems, cloud computing platforms, and MapReduce. In the context of our paper, massively parallel processing databases is relevant to address the real-time operation within tolerable elapsed time.

\subsection{Big Data Definition and its Features}
Big data is a data-driven cognitive approach; it perceives the world through data---it works out the statistical correlations indicated by high-dimensional parameters using a non-parameter model.
Currently, there exists no standardized definition for big data. This paper gives a mathematical definition below as our past work \mbox{\cite{qiu2013bookcogsen, qiu2014Intial70N, qiu2014MIMO,qiu2012bookcogpp}}.

\begin{itemize}
\item For each sampling time, data of \Data N-dimension are modeled as vectors, say \Belong {\Vx {i}}{\VRr {R}{N}};
\item The number of data samples, say \Data T, is large;
\item A function, \Data f(\DatassI {\Vx{}}{T}), is able to be defined;
\end{itemize}

One of the big data fundamental characteristics is huge volume of data represented by heterogeneous and diverse dimensions.
\Belong {\VX {}} {\VF CNT} is a natural model, formed by the data \DatassI {\Vx{}}{T}, to describe a large-scale system or subsystem \cite{qiu2014foundation}; \VX {} is a non-parameter model formed almost based on a minimum hypothesis.
While the size (rows $N$ for dimensions number; columns $T$ for samples number) increases, so do the complexity and the relationship underneath the data. Whereas,
when the size are sufficiently large, some unique phenomena, such as concentration of spectral measure \cite{qiu2015smart}, will occur.

\subsection{Random Matrix Theory}
Random matrix theory (RMT) has emerged as a particularly useful framework for many theoretical questions, especially for those concerning multivariate data.
There are two frameworks for RMT: one assumes the asymptotic convergence, and the dimensions are infinite; the other one is the non-asymptotic solution, assuming finite matrix size. For the asymptotic results, in theory, we require the infinite size of the matrix, which is infeasible in practical world. However, the results are remarkably accurate, even for relatively moderate matrix sizes such as tens. This is the very reason why this random matrix model is penetrating so many areas from financial engineering to wireless network. Our initial motivation for this model was from large-scale wireless network. The new trends for RMT are (1) finite matrix and (2) non-Gaussian matrix entries.

\subsubsection{Marchenko-Pastur Law (M-P Law)}
{\Text{\\}}

 M-P Law describes the asymptotic behavior of singular values of large rectangular random matrices. Let \Equ {\VX{}}{\{  \Vsx{i,j}\}}  be a \Muls NT random matrix whose entries, with the mean \Data {\Equs {\Mu{\Vsx{}}}{0}} and the variance \Data {\Sigg{\Vsx{}}\!<\!\INF}, are independent identically distributed (i.i.d.). As \Trend {N,T}{\INF} with the ratio \Belong{\Equ {c}{N/T}}{(0,1]}, the empirical spectrum density (ESD) of the corresponding sample covariance matrix \Belong {\Equs {\Vector S}{\Div 1 N\VX{}\Her{\VX{}}}}{\VF CNN} converges to the distribution of M-P Law \cite{qiu2012bookcogpp,marvcenko1967distribution} with density function

\FuncCC {eq:MPLaw}
{\Funcfzx {f}{\text{ESD}}{\Vlambda{\Vector S}}}
{ \Div{1}{2\pi{}\lambda c{\sigma^2}}\Sqrt{ (b\!-\!\lambda)(\lambda\!-\!a)} }{{\Text {, }} a\LE \lambda \LE b }
{0}  {\Text {, otherwise}}
where \Equs a{\sigma^2(1-\Sqrt c)^2}, \Equs b{\sigma^2(1+\Sqrt c)^2}.

\subsubsection{Kernel Density Estimation (KDE)}
{\Text{\\}}

A nonparametric estimate \cite{pan2011universality} of the ESD of the sample covariance matrix \Belong {\Vector S}{\VF CNN} is used
\begin{equation}
\Equ {\Funcfzx{f}{\text{ESD}}{\Vlambda{\Vector S}}}{\Div 1{Nh}\Sum{i=1}{N}{K(\Div{  \Data{x-\Vlambda {\Vector{S},i}}}{h})} }
\label{eq:fKDE}
\end{equation}
where \Vlambda {\Vector{S},i} (\Data {\Equs{i}{1,2,\cdots,N}}) are the eigenvalues of \Vector S, and $K(\cdot)$ is the kernel function for bandwidth parameter \Data h.

\subsubsection{Ring Law}
{\Text{\\}}

 Ring Law for (large) non-Hermitian matrices is one of the most remarkable developments in the modern probability  \cite{guionnet2009single, benaych2013outliers}. Except for our previous work \cite{qiu2013bookcogsen,qiu2014Intial70N,qiu2014MIMO,qiu2015smart}, little research has been done to leverage this new tool in the context of modeling massive datasets. This mathematical structure is general enough to model many unprecedented problems.

Consider the product  of $L$ non-Hermitian random matrices \Data {\VZ{}=\Prod{i=1}{L} \VX {u,i}}, where \Belong{\VX u}{\VF CNN} is the singular value equivalent \cite{ipsen2014weak} of the rectangular non-Hermitian random matrix \Belong{ \VTX{}}{\VF CNT},
whose entries are i.i.d. variables with the mean \Equs {\Mu{\Tdata{x}}}{0}  and  the variance \Equs {\Sigg{\Tdata{x}}}{1}. This product \VZ{} allows us to study the streaming datasets generated as a function of both space and time. The basic target of this study is the prospective architecture and its effectiveness. For simplification, we set \Data {L} to one in this paper and need not to discuss more on \Data {L}.
Beyond that, the product \VZ{}, by a transform which make the variance to \Data {1/N},  can be converted to \VTZ {} (i.e. \Equs {\Sigg{\VTsz{}}}{1/N}). Thus, the ESD of \VTZ {} converges almost surely to the limit given by

\FuncCC {eq:Lambda}
{\Funcfzx {f}{\text{ESD}}{\Vlambda {\VTZ {}}}}
{ \Div{1}{\pi{}c\Data {L}}{\Power{\Abs{\lambda}}{(2/\Data {L}-2)}}}  {{\Text {, }} \Power {(1-c)}{\Data {L}/2} \LE \Abs{\lambda} \LE \Data 1   }
{0}   {\Text {, otherwise}}
as \Trend {N,T}{\INF} with the ratio \Equ {N/T}{\Belong {c} {(0,1]} }.

\begin{figure}[htb]
\centering
\subfloat[\Equs L1]{
\label{fig:L1}
\includegraphics[width=0.23\textwidth]{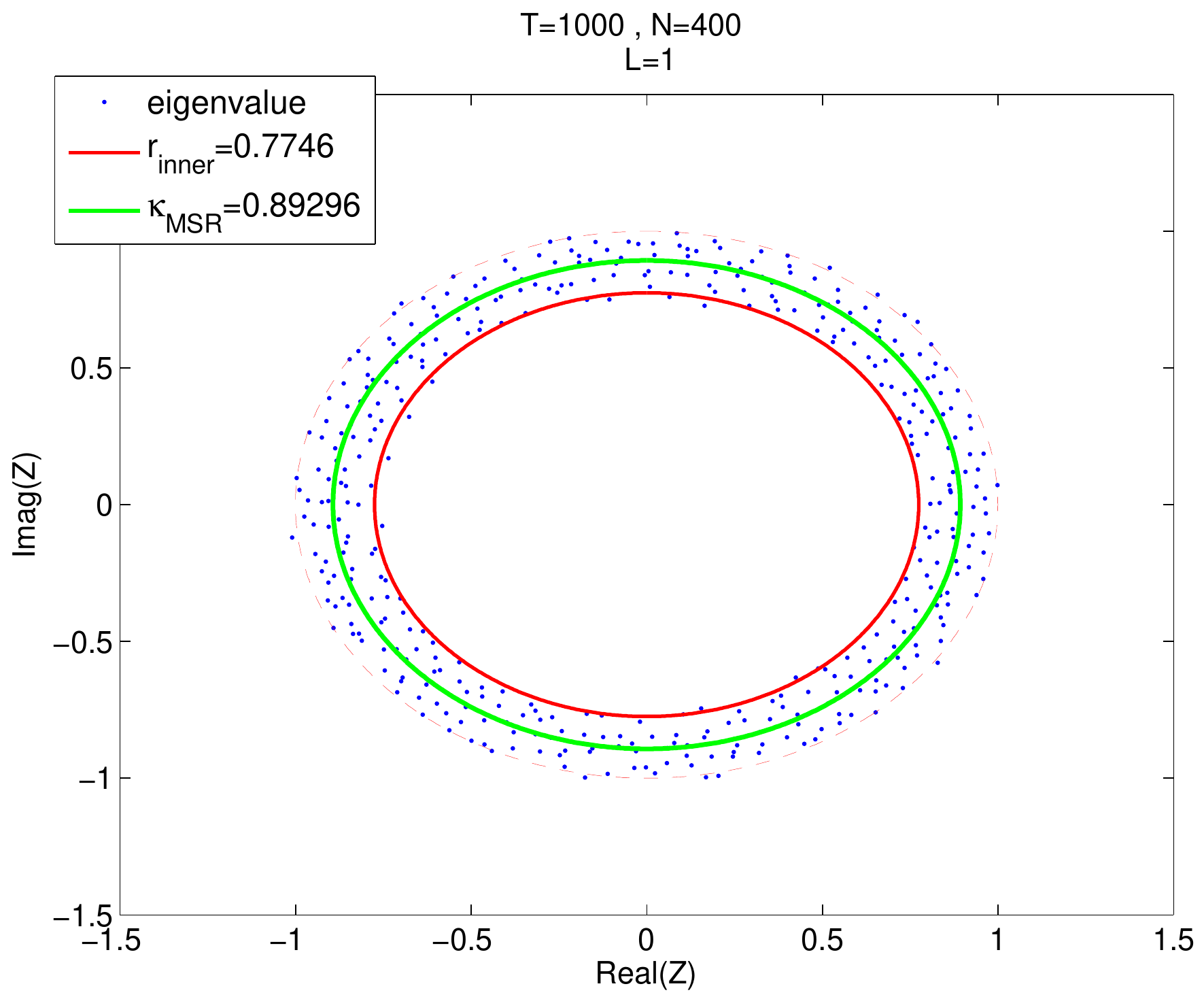}
\includegraphics[width=0.23\textwidth]{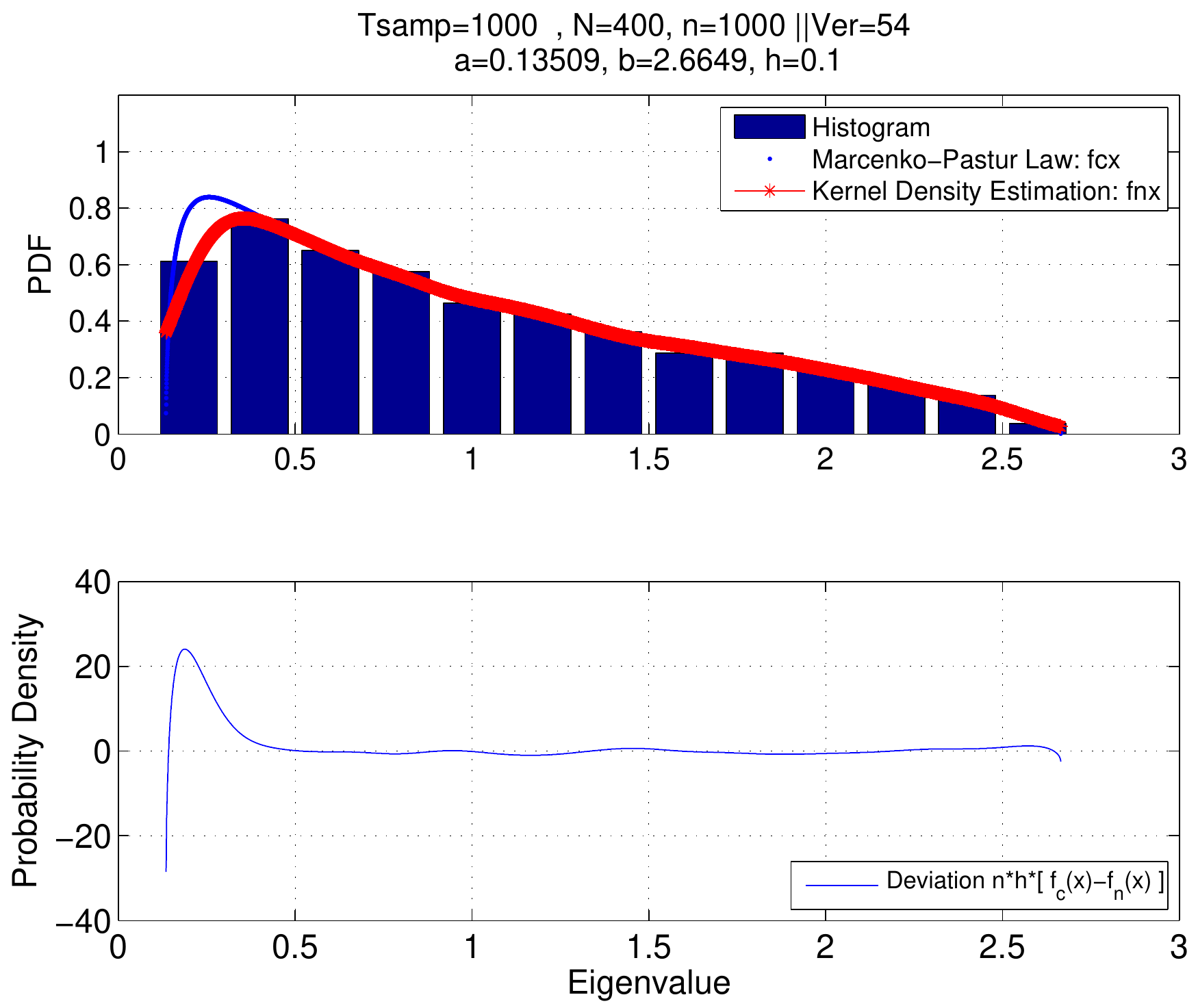}
}

\subfloat[\Equs L8]{
\label{fig:L8}
\includegraphics[width=0.23\textwidth]{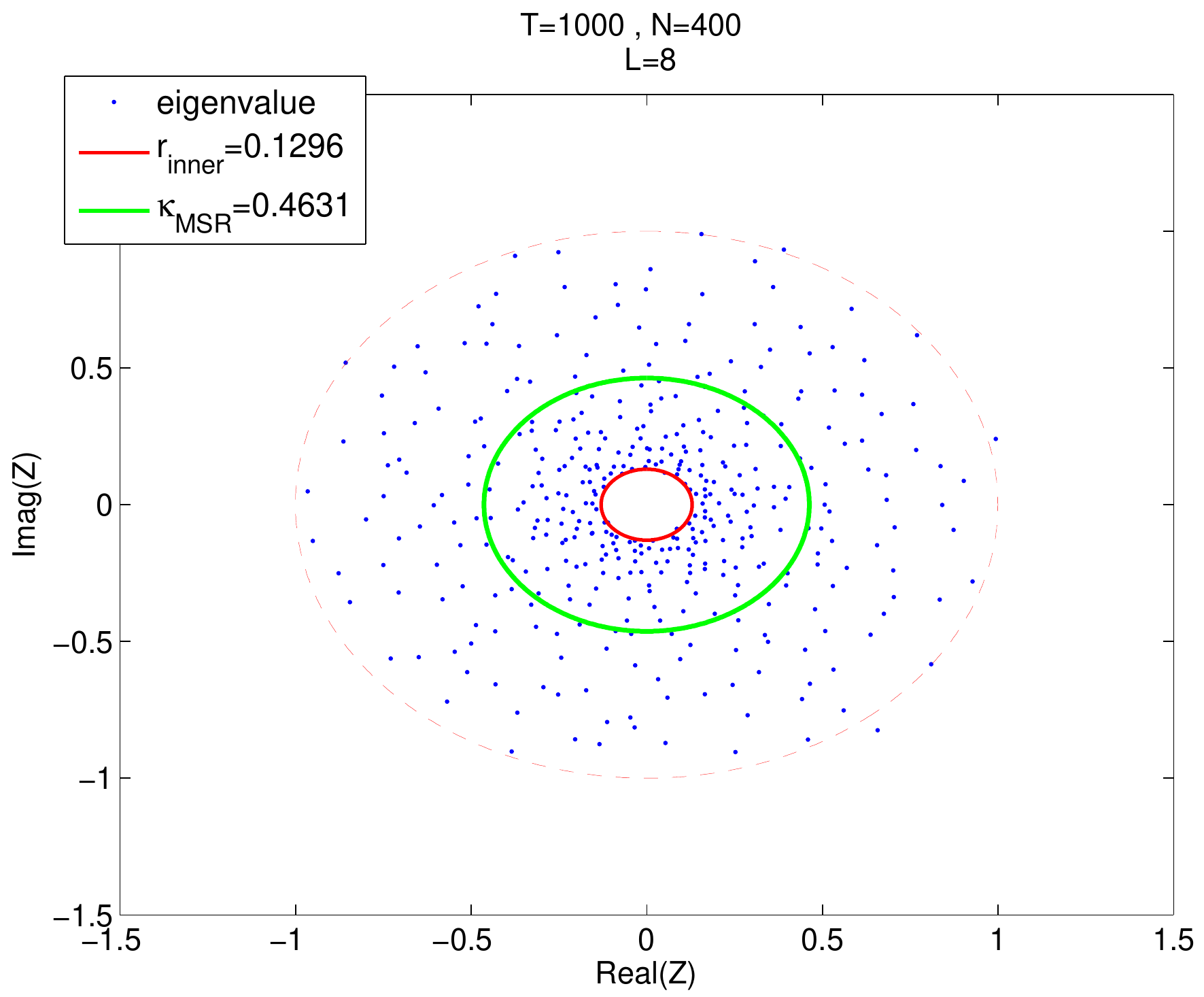}
\includegraphics[width=0.23\textwidth]{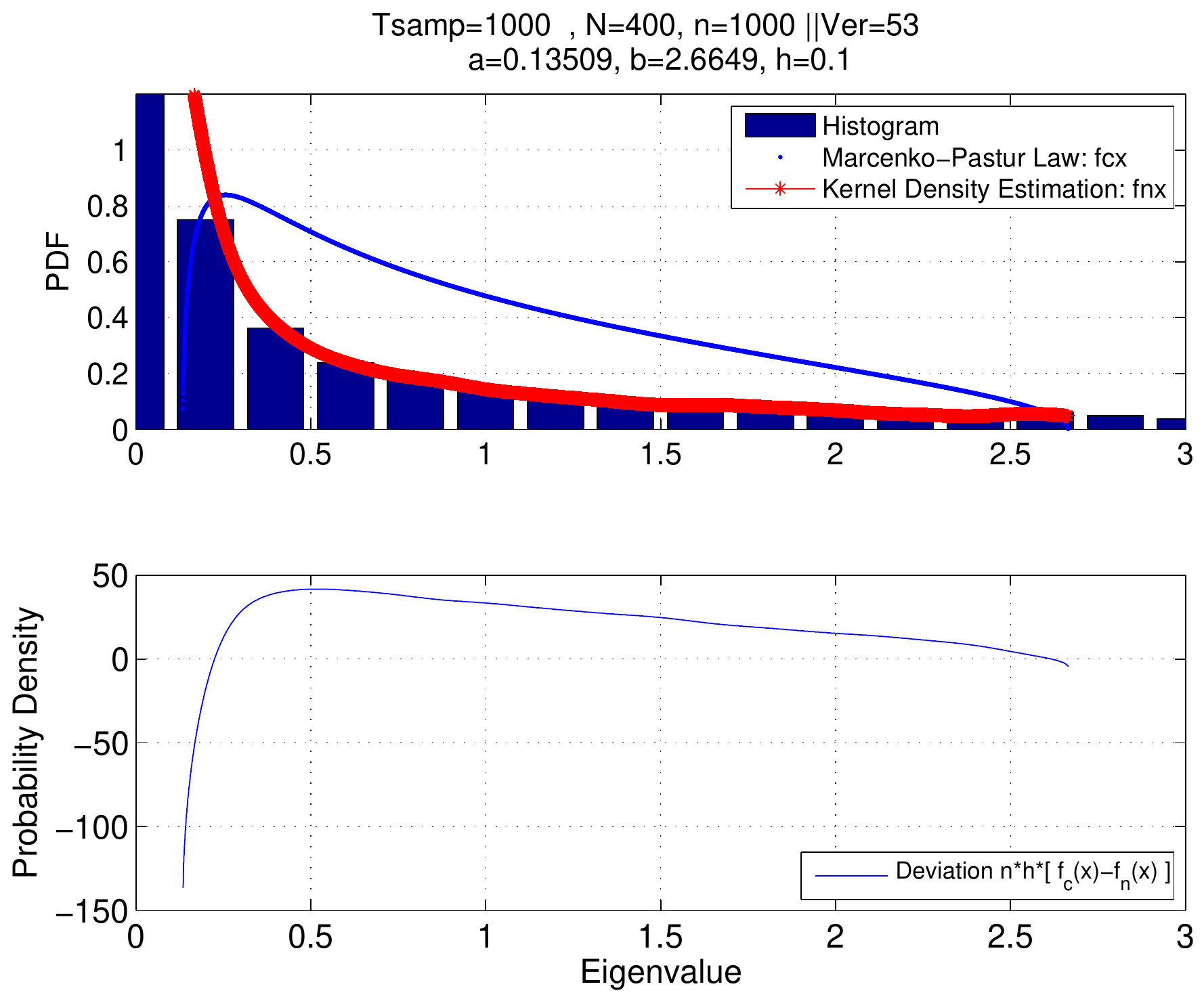}
}

\caption{Ring Law and M-P Law for \Data {\VZ{}=\Prod{i=1}{L} \VX {u,i}}}
\label{fig:ringlaw}

\end{figure}

On the complex plane of eigenvalues, the inner circle radius  is \Power{(1-c)}{\Data {L}/2} and the outer circle radius is unity.
Especially, \Equs {\Equs {\Vector S}{\VTZ{}\Her{\VTZ{}}}}{\Div{1}{N}\VY{}\Her{\VY{}}} ({\Equs{\VY{}}{\Sqrt{N}\VTZ{}}}\Belong{}{\VF CNN}, \Equs {\Sigg{\VTsz{}}}{1/N}, \Equs {\Sigg{y}}{\Sigg{\Sqrt{N}\VTsz{}}}\Equs{}{1}) is able to be acquired, and \Vector {S} conforms to M-P Law.
Ring Law and M-P Law with \Data {L=1 } and \Data {L=8} are shown as Figure \ref{fig:ringlaw}.
Furthermore, we propose \VMSR{} (defined at the end of this section) to indicate the eigenvalues distribution of \VTZ{} and the data correlation as a statistical parameter. In Figure \ref{fig:ringlaw}, \VMSR{} is depicted in green line.

\subsection{Big Data Analysis}

Data processing in power systems is a typical big data challenge.
For a system, lots of measured data or simulated ones \VRsx{} are readily accessible; for a certain time  \Vti{}, they are arranged as a vector \VRx{\Vti{}}. As time goes by, vectors are acquired one by one and a sheet is naturally formed as dataset to map the system. Any section in the dataset, according to our intend, is available as a raw data source \VOG{\VRx{}} for further analyses. Thus, \VOG{\VRx{}} consists of sample vectors on a series of times, which are denoted as \DatasII{\VRx{}}{t}{i}; at any time \Vti, the vector \VRx{\Vti{}}  consists of sample data  in various dimensions, which are denoted as \DatasII{\VRsx{}}{\Vti,n}{j} . The upper limit of dimensions \Data n (i.e. max \Data j) is subject to the variety of the data at a single sampling time, and the upper limit of time length \Data t (i.e. max \Data i) is subject to the volume of the database and generally big enough.

For the raw data source \VOG{\VRx{}}, we can focus on any data area (e.g., \Data N rows, \Data T columns) as a split-window;  a raw matrix \Belong {\VRX{}}{\VF CNT} is naturally formed. Then, \VRX{} is converted to a normalized non-Hermitian matrix \Belong {\VTX{}}{\VF CNT}  row-by-row with following algorithm:
\FuncC {eq:StdMatrix}
{\small{
\VTsx{i,j}\!=\!\Muls{  (\VRsx{i,j}\!-\!\Udata {\VRx{i}}) } {( \Sig{\VTx{i}}/\Sig{\VRx{i}} )\! + \! \Udata {\VTx{i}}}
\,,\Data {1\!\LE{}\!i\LE{}\!N\!;1\!\LE{}\!j\!\LE{}\!T}}
}
where \Equs {\VRx{i}}{(\DatasNTt {\VRsx{}}{i}{T})}  and \Equs {\Udata {\VTx{i}}}{0}, \Equs {\Sigg {\VTx{i}}}{1}.

The  matrix \Belong {\VX {u}} {\VF CNN} is introduced as the singular value equivalent \cite {qiu2015smart} of  \Belong {\VTX{}} {\VF CNT} by
\FuncC {eq:Xu}
{\Equ {\VX u}   {\Sqrt{\VTX{}{\Her{\VTX{}}}}\Vector U }}
where \Belong {\Vector U} {\VF CNN} is a Haar unitary matrix, \Equu {\VX{u}\Her{\VX{u}}} {\VTX{}{\Her{\VTX{}}}}.

For \Data L arbitrarily assigned  independent non-Hermitian matrices  \VRX{i} (\Equs{i}{1,\!\cdots\!,L}) in the raw data source \VOG{\VRx{}}, the matrices product \Equ {\VZ {}}{ \Prod{i=1}{L}{\VX {u,i}}}\Belong{}{\VF CNN}  is able to be acquired. Then, \VTZ{} is calculated row-by-row with following formula:

\FuncC {eq:Z}{
{\Equ {\VTz{i}} {\Vz{i}/({\Sqrt N}\Sig{\Vz{i}})}}
\quad ,\Data {1\LE{}i\LE{}N}
}
where \Equs {\Vz{i}}{(\DatasNTt {\Vsz{}}{i\!,}{N}) }  , \Equs {\VTz{i}}{(\DatasNTt {\VTsz {}}{i\!,}{N}) }.

Furthermore, for the radius of all eigenvalue of \VTZ{} on the complex plane, we calculate its mean value and denote it as \VMSR{} (i.e. \Equs {\VMSR{}}{\Udata {\Vector r_{\Vlambda{\VTZ{}}}}}). In general, we conduct high-dimensional analysis, only with its dataset, to reveal the properties of a power system. \VTZ{} and its ESD are analyzed based on the newly developed Ring Law, and the high-dimensional statistic \VMSR{} is calculated and visualized as an indicator. In addition, the corresponding sample covariance matrix  \Equs {\Vector S}{\VTZ{}\Her{\VTZ{}}}   is calculated for the comparison among the results of its histogram, KDE, and  M-P Law.

\section{A Big Data Architecture for Smart Grids and its Advantages}

\subsection{Big Data Architecture for Smart Grids}
The frequently used notations are shown in Table \ref{tab:Notations}.

\begin{table}[htbp]
\caption{Notations for Architecture and Case Studies}
\label{tab:Notations}
\centering

\begin{tabularx}{0.48\textwidth} { l !{\color{black}\vrule width1pt} p{6.8cm} } 

\toprule[1.5pt]
\hline
\textit{Notations} & \textit{Means}\\

\Xhline{1pt}

\VAT{} & time area to focus on the data split-window \\
\VIndex A{\Text{Node}} &	node area to focus on the data split-window\\
\Data {t} &	time: \Index t{0} for current time, \VtS{} for sampling time\\

\VPbus{n} &	power demand of load at bus-n\\
\VPbusmax{n} &	critical point on \Cur{P_{\!{ower}}}{V_{oltage}} curve for bus-n\\
\Vgam{\Text {Acc}} & addition factor for load fluctuations of the grid\\
\Vgam{\Text {Mul}} & multiplication factor for load fluctuations of the grid\\

\toprule[1pt]

\end{tabularx}

\end{table}

The designed architecture is illustrated as Fig. \ref{fig:architecture}.

\Figf {fig:architecture}{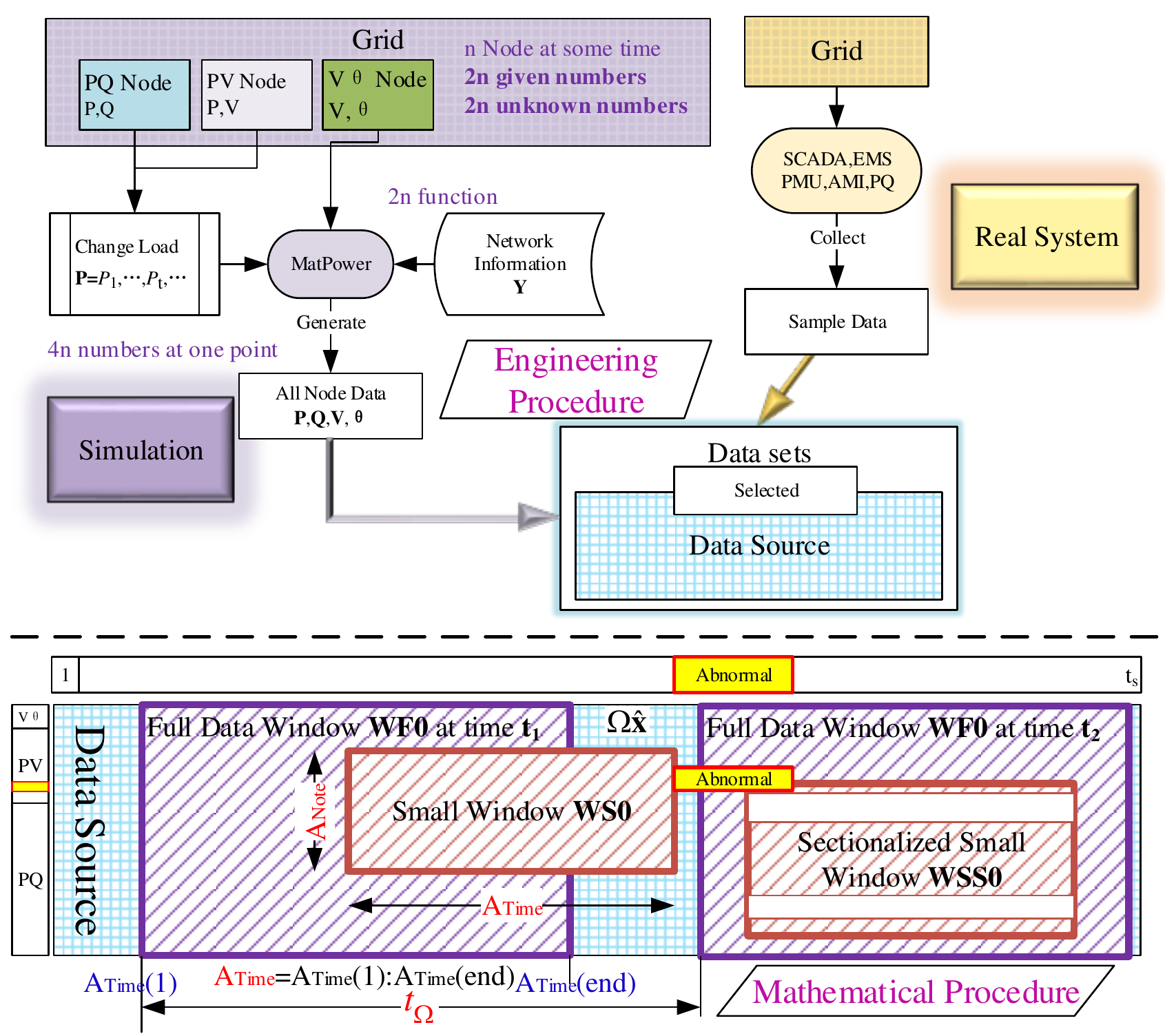}{The designed big data architecture for smart grids. The part above the dot line illustrates an engineering procedure for big data modeling, during which the raw data source \VOG {\VRx{}} are formed to map the physical system. The other part below the dot line illustrates a mathematical procedure for big data analysis. It is fully independent of engineering parameters, and during which the analyses are extracted from data source \VOG {\VRx{}}.}{htbp}

It consists of two independent procedures to connect smart grids and big data---big data modeling as an engineering procedure, following by big data analysis as a mathematical procedure. During the engineering procedure, the raw data source \VOG{\VRx{}} is acquired as described in Section II; during the mathematical procedure, following steps are conducted:

\begin{table}[!h]
\centering

\begin{tabular*}{8.8cm} {lp{8.6cm}}
\toprule[1.5pt]
\textbf {Steps of Mathematical Procedure} \\
\toprule[0.5pt]
1) Set the initial parameters\\
\qquad{}1a) Set \VIndex A{\Text{Time0}} and \VIndex A{\Text{Node0}} to focus on the first data window\\
\qquad{}1b) Set \Index t {\VOG{}} and \Equs k{0} to slide the moving split-window (MSW)\\
2) Focus on correspond window to form \VRX{} (\Equs{\VIndex A{\Text{Time}}}{\VIndex A{\Text{Time0}}+k})\\
3) Calculate \VTX{}, \VX{u}, \VZ{}, \VTZ{}, \Vector S\\
4) Calculate the eigenvalues \Vlambda {\VTZ{}},\Vlambda {\Vector S}\\
5) Conduct ESD analysis and compare the result according to RMT\\
6) Calculate \VMSR{} with \Vlambda {\VTZ{}}\\
7) Visualize the results\\
8) Judge as times goes by:\\
\qquad{}8a) \Data{k<{\Index t{\VOG{}}}\Rightarrow{}k\!+\!+}; back to \textit{step 2)}\\
\qquad{}8b) \Data{k\GE{\Index t{\VOG{}}}\Rightarrow{}\Text { END}})\\

\toprule[1pt]

\end{tabular*}

\end{table}

Especially, in step \textit{2) Focus on the data window}, we are able to conduct 1) real-time analysis: focusing on the real-time data window whose last edge of the sampling time area is current time (i.e. \Equs {\VIndex A{\Text{Time}}\Text{(end)}}{ \Index t0}); and 2) block-calculation for decoupling interconnected system: focusing on a smaller window consisting of data only in designated dimensions, but not in all dimensions.
Besides, as the split-window, in a fixed size, slides across the data source \VOG {\VRx{}} with \Index t {\VOG{}} and k set in \textit{1b)}, a series of  \VMSR{} is got for further research and visualization.

\subsection{Advantages in Data Processing}
\subsubsection{Algorithm}
{\Text{\\}}

This architecture analyzes data in high-dimension as illustrated by the solid purple lines in Fig. \ref{fig:procedure}. It is a universal procedure with 4 steps as follows:

\begin{table}[!h]
\centering

\begin{tabular*}{8.8cm} {p{8.6cm}}
\toprule[1.5pt]
\textbf {Steps of Data Management for G3} \\
\toprule[0.5pt]
1) Form standard random matrices \VTX{}\\
2) Acquire \VTZ{}, \Vector S by variables transforming (\Data{\VTX{}\rightarrow\VX{u}\rightarrow\VZ{}\rightarrow\VTZ{}\rightarrow\Vector S})\\
3)  Conduct high-dimensional analysis based on RMT\\
4) Conduct engineering interpretations\\

\toprule[1pt]
\end{tabular*}
\end{table}

On the other hand, the procedure of traditional data processing algorithms, in most cases, relies highly on specific simplifications and assumptions to build models. Taking genetic algorithm for an example, two steps are essential to achieve the result. One is to transcode the engineering variables to gene as the input of the gene model. It is a subjective selection procedure for the specific roles in engineering system, and only a few variables can be taken into account in final model. The other one is to perform the genetic algorithm through  operations of selection, crossover, and mutation. Many problems, such as improper settings of the population size, of the crossover probabilities, or of the mutation probabilities, will inevitably make the result worse.

Compared to traditional algorithms, big data analysis, driven by data, enables us to analyze the interrelation and interaction among all the elements in system, seen as correlations indicated by high-dimensional parameters.  Using a pure mathematical procedure without physical models and hypotheses, big data analysis is easier in logic and faster in speed. Moreover, except step \textit{4): Conduct Engineering Interpretation}, the whole procedure is objective without introducing or accumulating the systematic errors; moreover, the accidental errors can be eliminated either with the matrix size growing, or by repletion test and parallel computing due to the independence of the algorithm.

\subsubsection{Distributed Calculation for Interconnected Grids}
{\Text{\\}}

Smart grids operation are featured with autonomous sources and decentralized controls. Due to the potential transmission cost and privacy concerns, Aggregating distributed data sources to a centralized site for mining is systematically prohibitive.
On the other hand, although we are able to carry out model-based mining at each distributed site, the decoupling procedure for connected sources is high related to simplifications and assumptions. Accordingly the result are often barely satisfied and leads to biased views and decisions.

Large random matrices provide a natural and universal data-driven solution. For a specific zone-dividing interconnected system, each site is able to form a small matrix only with its own data. In this way, the integrated matrix can be divided into blocks for distributed calculation and vice versa. For the overall system, we can conduct high-dimensional analysis by integrating the regional matrices, or even by processing a few regional high-dimensional parameters. The mathematical  foundation is kept invariant as RMT; the scalability, however, depends on our intention. This architecture, decoupling the systems in the form of statistical matrices or high-dimensional parameters instead of models, is practical for real large-scale interconnected grids. The distributed calculation is a deep research, and in this paper we just provide a relative simple applied example---\textit{case 4} in the next section \textit{Five Case Studies}.

\subsection{Advantages in Management Mode}

Some brief but novel introductions and analyses about the development of power grids, mainly under the  perspective of data managing mode and information communication technology (ICT), are given as the related background for applying big data into smart grids. Meanwhile, new situations and challenges, and advanced management mode for future grids are further discussed from the perspective. Especially, it is group-work mode, in our opinion, that breaks through the regional limitation for energy flows and data flows. As a result, some data-driven functions, e.g., comparative analysis, and distributed calculation, are able to be carried out.

Generally, the power grid’s evolutions are summarized as three generations---G1, G2, and G3 \cite{zhou2013review}. Their own network structures are depicted in Fig. \ref{fig:G3s} \cite{mei2014theevolution}. Meanwhile, their data flows and energy flows, as well as corresponding data management systems and work modes, are quite different \cite{he2014power}, which are shown in Fig. \ref{fig:flow} and Fig. \ref{fig:procedure}, respectively. In the following discussions, we will come to a conclusion that the group-work mode is the precondition for data-driven analysis, and the trend for smart grids.

\Figf {fig:flow}{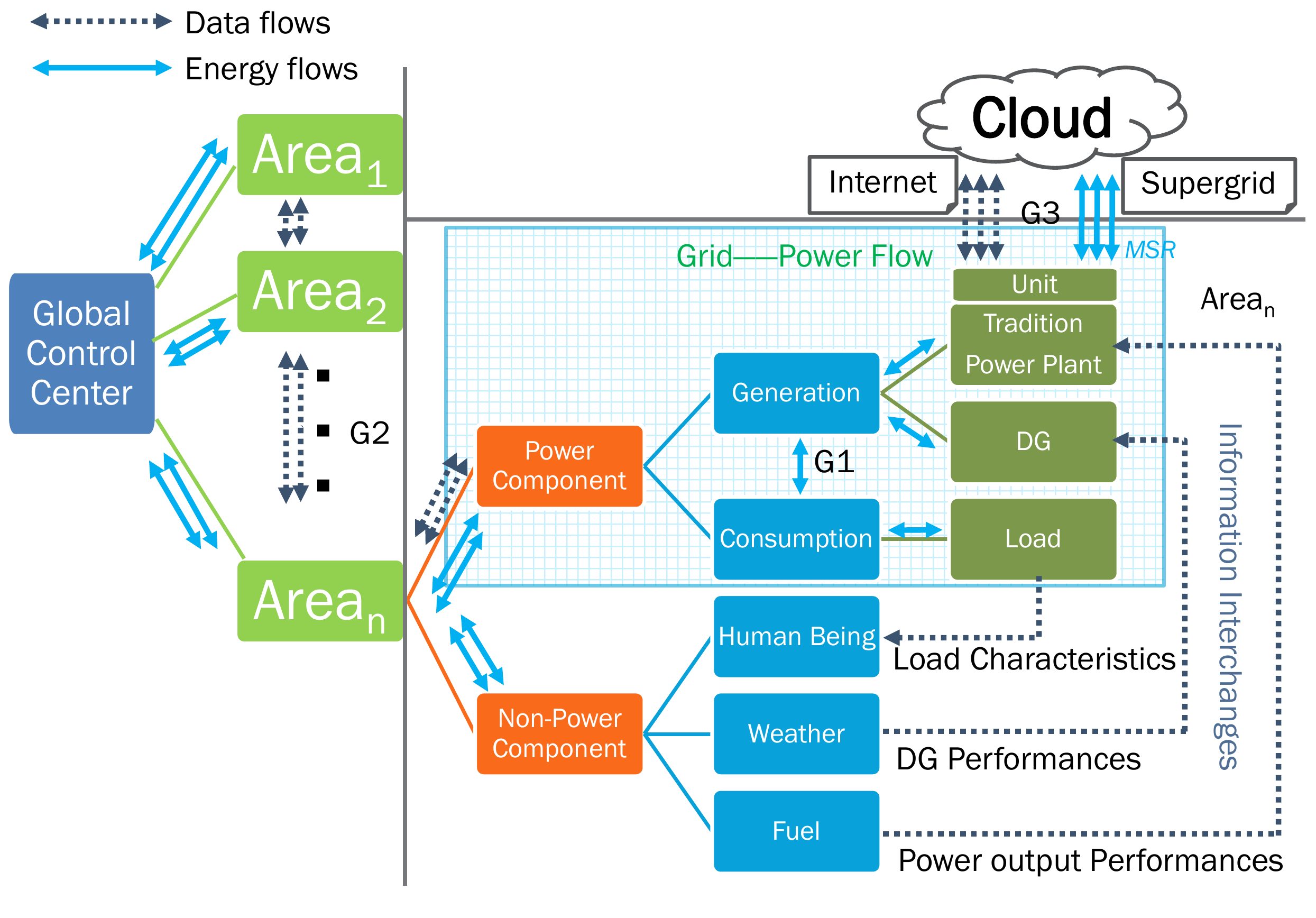}{Data flows and energy flows for three generations of power systems. The single lines, double lines, and triple lines indicate the flows of G1, G2, and G3, respectively.}{htbp}

\Figfp {fig:procedure}{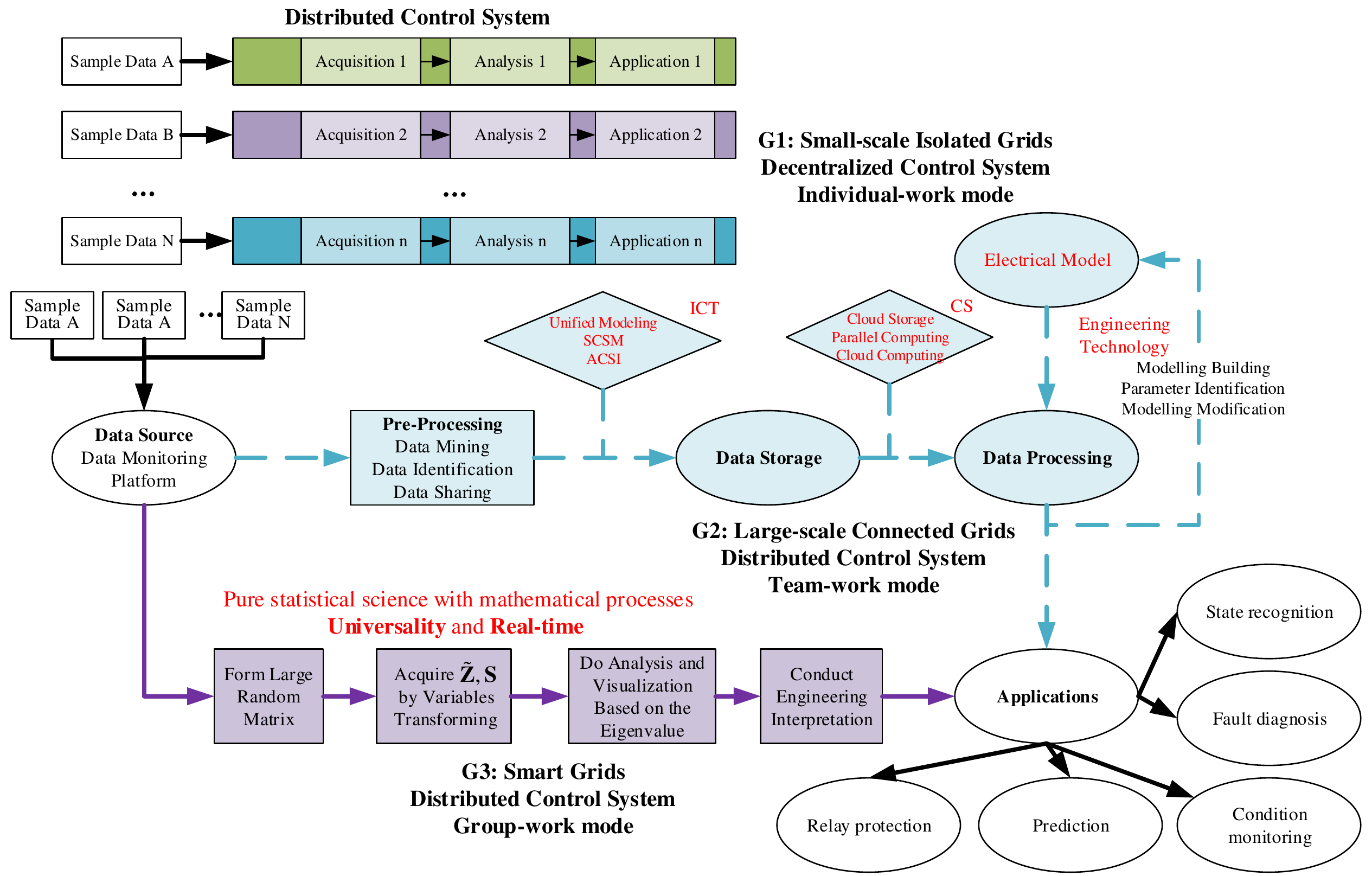}{Data management systems and work modes for three generations of power systems. The above, middle, and below parts indicate the data management systems and the work modes of G1, G2, and G3, respectively. For G1, each grid works independently. For G2, global and local control centers are operating under the team-work mode. For G3, the group-work mode breaks through the regional limitation for energy and data flows and has a better performance.}{htbp}


\subsubsection{G1: Small-scale Isolated Grids}
{\Text{\\}}

G1 was developed from around 1900 to 1950, featured by small-scale isolated grids. For G1, components interchange energy and data within the isolated grid to keep stable. The components are fully controlled by decentralized control system and operating under individual-work mode. It means that each apparatus collects designated data, and makes corresponding decisions only with its own application, just as shown at the above part of Fig. \ref{fig:G3sa}. The individual-work mode works with an easy logic and little information communication. Whereas, it means few advanced functions and inefficient utilization for resources. It is only suitable for small grids.

\subsubsection{G2: Large-scale Interconnected Grids}
{\Text{\\}}

G2 was developed from about 1960 to 2000, featured by zone-dividing large-scale interconnected grids. For G2, utilities interchange energy and data within the adjacent ones. The components are dispatched by control center and operating under team-work mode. The regional team leaders, likes local dispatching centers, substations, or microgrid control centers, aggregate their own team-members (i.e. components in the region) into a standard black-box model. These standard models will be further aggregated by the global control center  for  the control or prediction purposes. The two aggregations above are achieved by four steps, which are data monitoring, data pre-processing, data storage, and data processing, respectively. However, lots of engineering technologies or sciences are essential as the foundations---Cognitive Radio Wireless Network \cite{qiu2011papercog,qiu2012efficient,qiu2012scheduling},  Specific Communication Service Mapping (SCSM) as Information and Communication Technology (ICT) \cite{IEC2004iec}, Cloud Storage, Parallel Computing  as Computer Science (CS), and Modelling Building, Parameter Identification as Mathematical Modeling \cite{he2013research,he2014impact}. The description above are illustrated by dotted blue lines in Fig. \ref{fig:procedure}. In general, the team-work mode conducts model-based analysis, and mainly concerns with the system stability rather than the individual benefit; it will not work well for smart grids with 4Vs data as described in \textit{Section I}.

\subsubsection{G3: Smart Grids}
{\Text{\\}}

The development of G3 was launched at the beginning of the 21st century; and for China, it is expected to be completed around 2050 \cite{zhou2013review}. Fig. \ref{fig:G3sc} shows that the clear-cut partitioning is no longer suitable for G3, as well as the team-work mode which is based on the regional leader. For G3, the control force of the regional center (if still exist) is greatly released by individual units. The high performance and self-control  individuals results in much more flexible flows, for both energy exchange and data communication, to improve utilization by sharing resources among the whole grid \cite{zhang2014economic}. Accordingly, the group-work mode is proposed. Under this mode, the individuals play a dominant part in the system under the authority of the global control centers \cite{he2014power}. VPPs \cite{ai2011multi-agent}, MMGs \cite{he2012research}, for instance, are typically G3 utilities. These group-work mode utilities provide a relaxed environment to benefit both the individuals and the grids: the former (i.e. individuals), driven by their own interests and characteristics, are able to create or join a relatively free group to benefit mutually from sharing their respective superior resources; meanwhile, these utilities are generally big and controllable enough to be good customers or managers to the latter (i.e. the smart grids).

\section{Five Case Studies}
For the following designed experiments, all the data are obtained in two scenarios: 1) Only white noises (i.e. benchmark, whose statistical results agree with M-P Law and Ring Law); 2) Signals plus noises. We treat small random loads fluctuations and sample errors as white noises, and sudden changes and faults as signals.

Case 1 to Case 4, based on Matpower, belongs to the field of power system stability and control. The grid is a standard IEEE 118-bus system with six partitions displayed as Fig. \ref{fig:IEEE118network}  \cite{ni2007new}. Detailed information about the test bed is referred to the \textit{case118.m} in \textit{Matpower package} and \textit{Matpower 4.1 User’s Manual} \cite{MATPOWER2011matpower}. Case 5, based on PSCAD/EMTDC, is about fault detection.

\subsection{Case 1: Observation from the Split-Window with Full Network and 500 Sample Points---\Equs {N}{118}, \Equs {T}{500}}

\begin{figure}[htbp]
\centering
\subfloat[\EtS{550}{s}]{
\label{fig:case1a}
\includegraphics[width=0.23\textwidth]{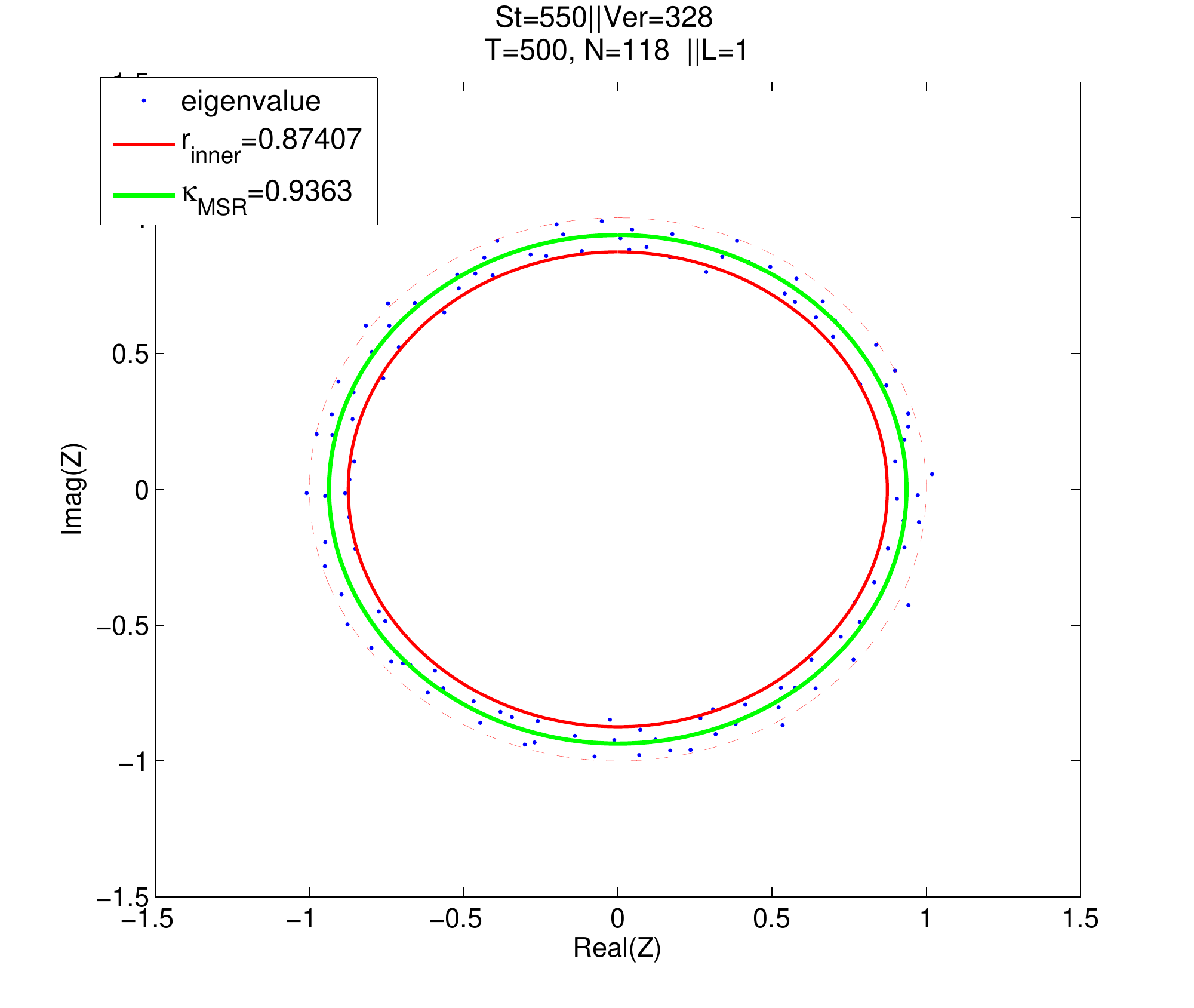}
\includegraphics[width=0.23\textwidth]{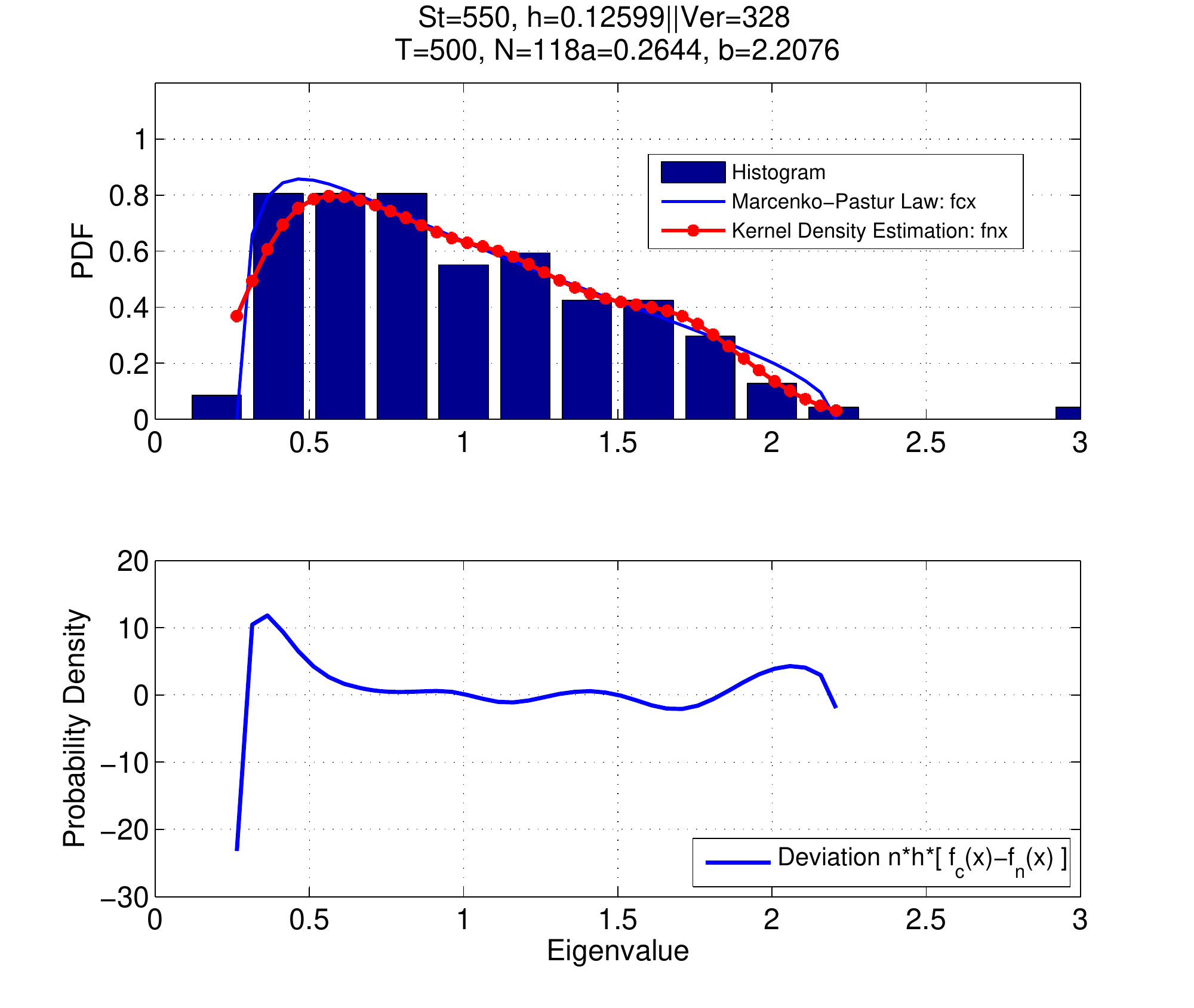}
}

\subfloat[\EtS{551}{s}]{
\label{fig:case1b}
\includegraphics[width=0.23\textwidth]{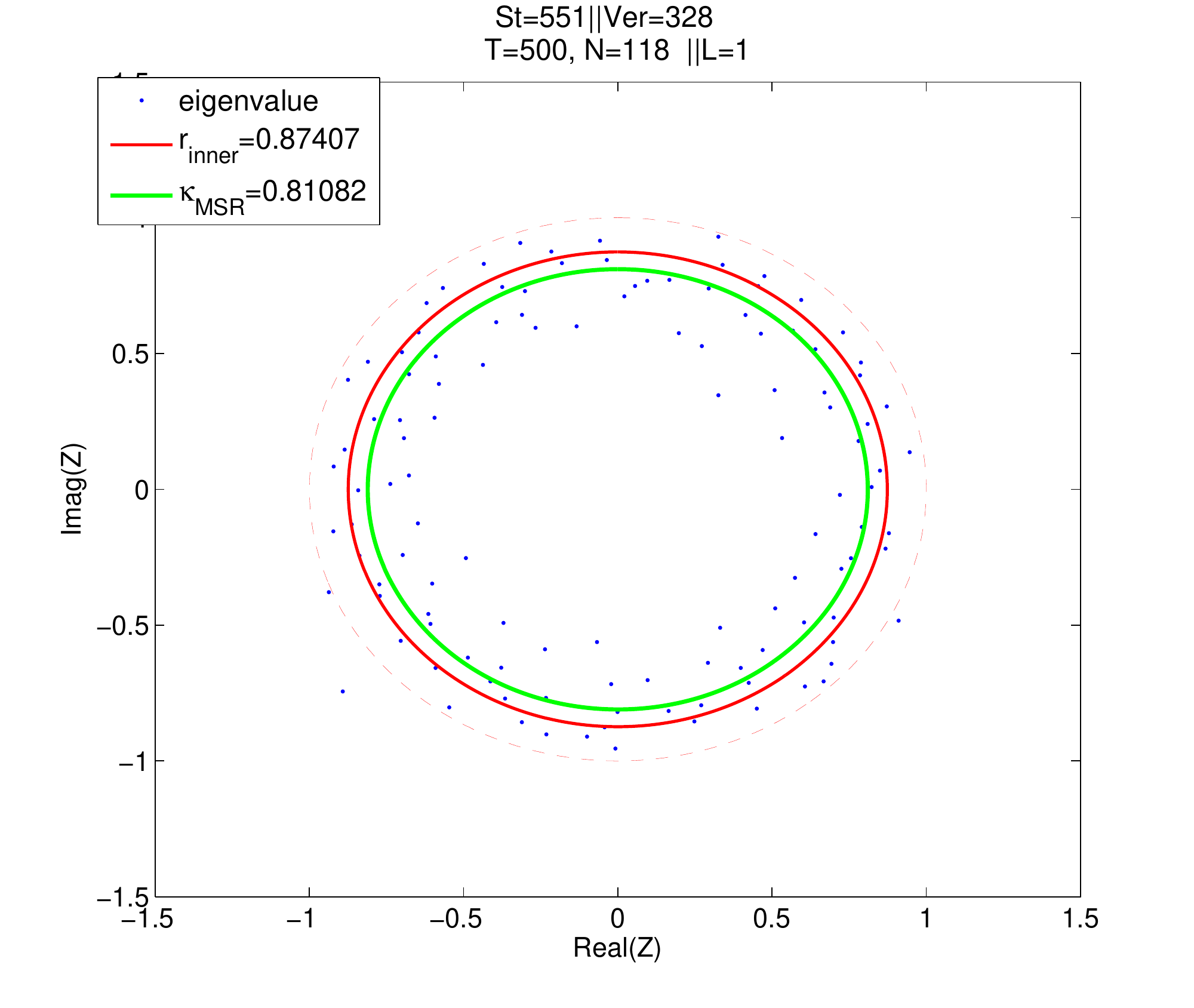}
\includegraphics[width=0.23\textwidth]{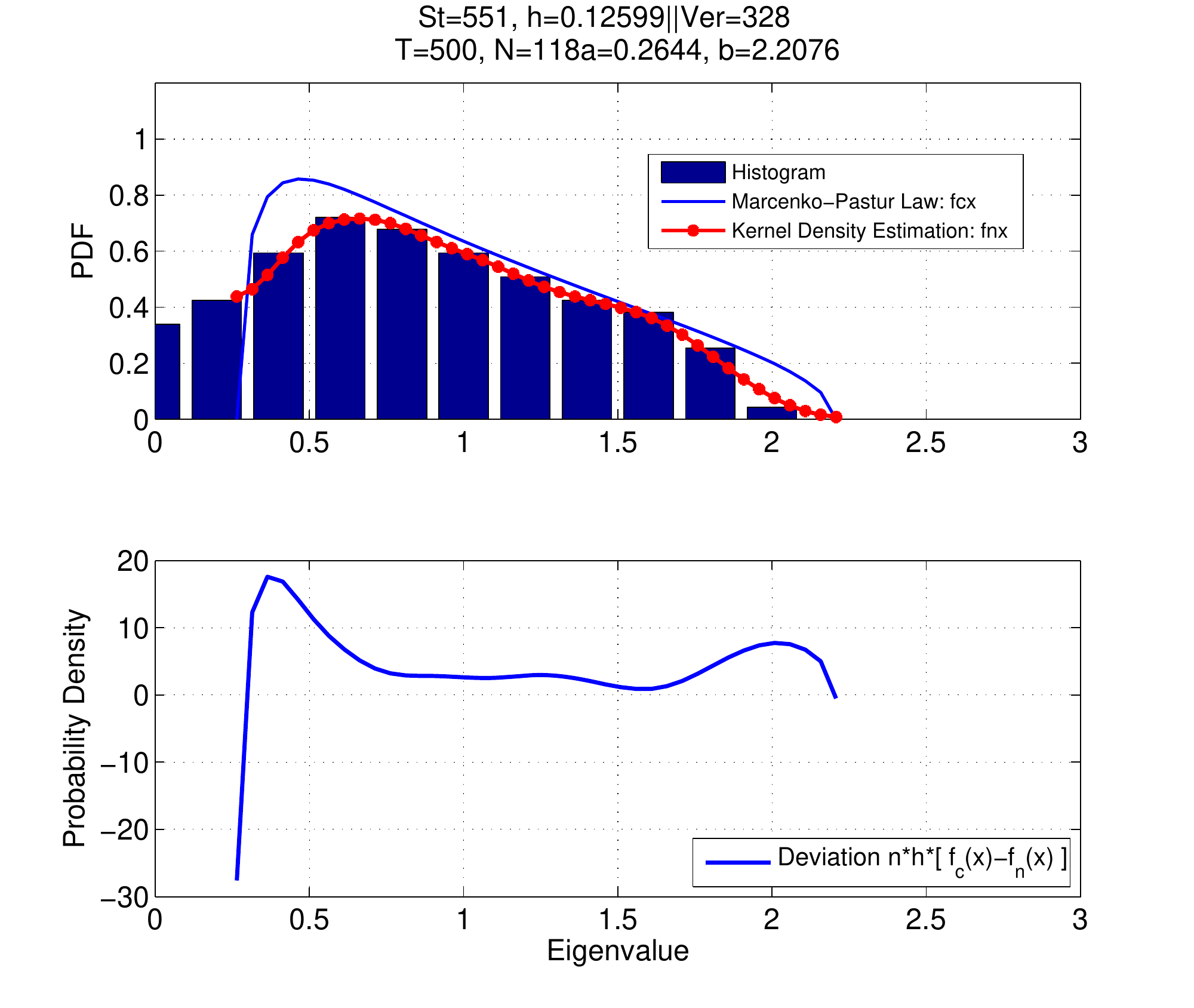}
}
\caption{Ring Law, and the Comparation among Histogram, KDE, and M-P Law at Critical Time for Case 1}
\label{fig:case1ab}

\includegraphics[width=0.48\textwidth]{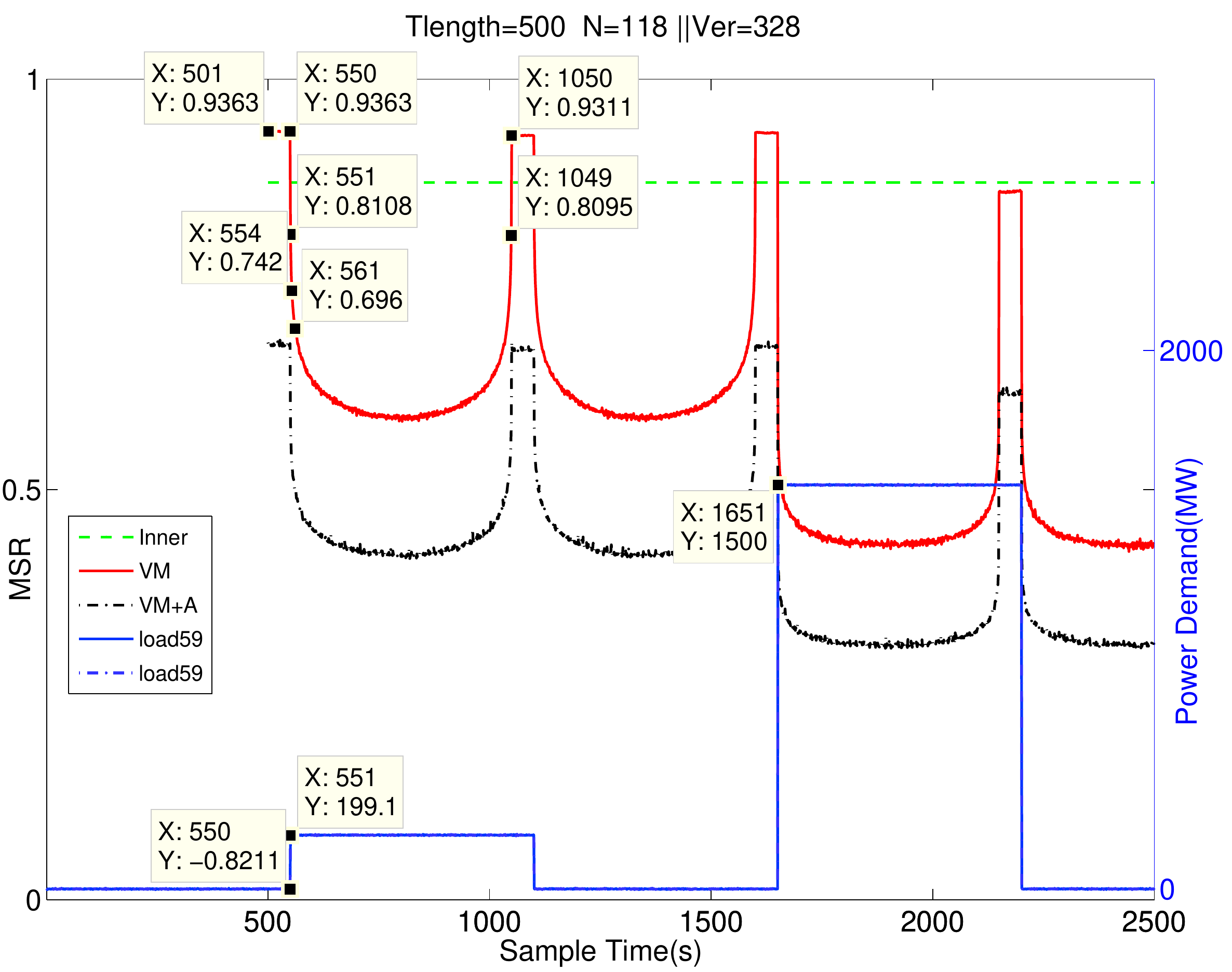}
\caption{MSRs on the Time Series  for Case 1}
\label{fig:case1c}
\end{figure}

Case 1 is a simple study to validate that the architecture is able to quickly detect the signals from the noises with the real-time data flow.
Let \Equs N{118}, \Equs T{500}, \Equss c{N/T}{0.236}. Thus each split-window has \Equss {n}{NT}{59,000} data. Table \ref{tab:case1es} shows the series of assumed events, and the accordingly RMT analysis results visualization at critical time and MSRs on the time series are depicted as Fig. \ref{fig:case1ab} and \ref{fig:case1c}, respectively.

\subsubsection{Sampling time \EtS{550}{s}, Time Area \EAT{51\!:\!550}{s}}
{\Text{\\}}

There are only small load fluctuations in this split-window, which means that white-noises play a dominant part. Then, we compare the histogram, KDE and M-P Law. Inspection of Fig. \ref{fig:case1a} indicates that, in a white-noises dominated system, the kernel density estimation (in red line) matches the histogram (in blue bar) very well. Moreover, the histogram curve and the KDE curve agree with M-P Law (in blue line).

\subsubsection{Sampling time \EtS{551}{s}, Time Area \EAT{52\!:\!551}{s}}
{\Text{\\}}

For this split-window, Fig. \ref{fig:case1b} shows that the eigenvalues of Ring Law collapse to the circle center, and both the histogram and KDE deviate from M-P Laws. It means that there are some signals, any deviation of the benchmark (white noises only), in the system. Indeed, just at time \Et {551}{s}, the \VPbus{59} suddenly changes from $0$ MW to $200$ MW somehow.

Fig. \ref{fig:case1c} depicts that the \VMSR{} change dramatically in a short time since \Et {550}{s}. As the length of time area \Equs {T}{500}, the step change as the signal occurring at time \Et {551}{s}, is included during all the sampling times from \EtS {551}{s} to \EtS {1049}{s}. It results in the deviation  ($0.9363, 0.8108, 0.742,\cdots$). However, at the sampling time \EtS {1050}{s}, when the time area \EAT{551\!:\!1050}{s}, the step signal is no longer exist, as well as the deviation of the histogram and KDE from M-P Law, and \VMSR{} is back to $0.9311$.

In addition, it is found that \VMSR{} for the data of $V$ (red line) which has definite physical meaning, and of $V+i\theta$ (black line) without any physical meaning, have the same trend. It indicates that MSR is a high-dimensional statistic, which is independent and robust to physical model in some way. The green line indicates the inner radius of the ring for the analyzing matrix, whose value only depends on the matrix size as formula \ref{eq:Lambda} in section II.

This case indicates that the presented statistic MSR is sensitive to signal. Meanwhile, there are some inherent relations for the data in different dimensions. It means that real-time analysis for event detection can be carried out with less kinds of data.

\subsection{Case 2: Observation from the Smaller Split-Window with Full Network and 240 Sample Points---\Equs {N}{118}, \Equs {T}{240}}

\begin{figure}[htbp]
\centering
\includegraphics[width=0.48\textwidth]{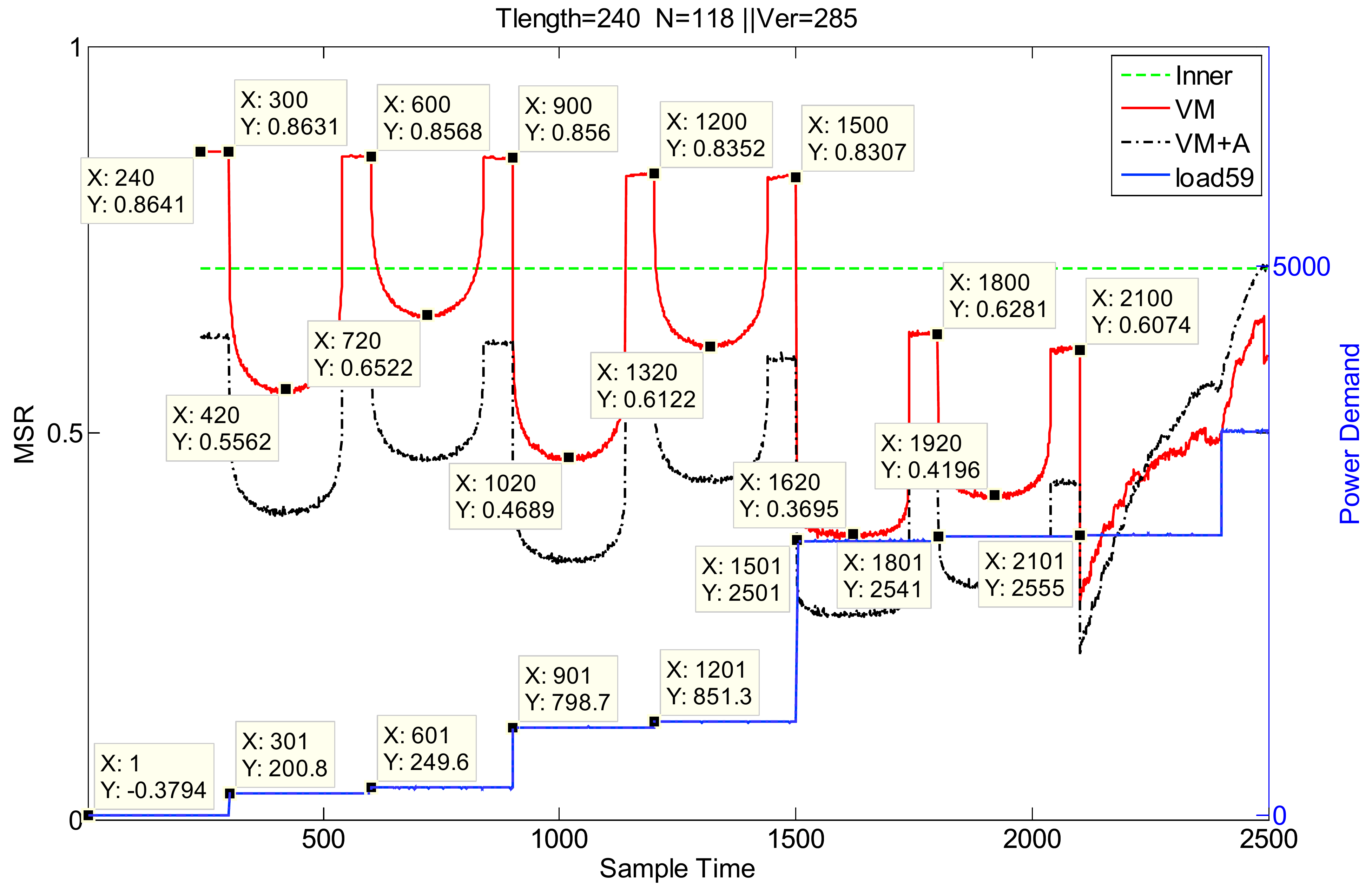}
\caption{MSRs on the Time Series  for Case 2}
\label{fig:case2}
\end{figure}

Case 2 is similar to the previous one except that \Data {T} decreases to 240 s, and the events are rearranged. In this case, we try to find the qualitative relationships between the MSR and the quantitative parameters of system performances. The series of assumed events and the \Cur{\VMSR}{t} curve are depicted as Table \ref{tab:case2} and Figure \ref{fig:case2}, respectively.

a) During once step-change, there is a negative correlation between the min value of  MSR (i.e. \VMSRmin{}) and the step-change value of \VPbus{59} (i.e. \VDelta {\VPbus{59}}):

\begin{table}[!h]  
\centering
\begin{tabular*}{8.8cm} { l  !{\color{black}\vrule width1pt} >{$}l<{$}  >{$}l<{$} >{$}l<{$} >{$}l<{$} >{$}l<{$}  }

\Xhline{1.5pt}

\VDelta {\VPbus{59}}& 200   &  50 &  550 & 1650 & \cdots\\
\VPbus{59}  & 0\!\rightarrow{}\!200 &  200\!\rightarrow{}\!250 &  250\!\rightarrow{}\!800 &
850\!\rightarrow{}\!2500 &\cdots \\
\VMSRmin{} & 0.5562 & 0.6522 & 0.4689 & 0.3695 & \cdots\\

\Xhline{1pt}

\end{tabular*}
\end{table}

b) When \VPbus{59} is steady at a higher level, \VMSR{} is steady at a lower level:

\begin{table}[!h]  
\centering
\begin{tabular*}{8.8cm} { l  !{\color{black}\vrule width1pt}>{$}l<{$} >{$}l<{$} >{$}l<{$} >{$}l<{$}>{$}l<{$}>{$}l<{$}  }  

\Xhline{1.5pt}

\VPbus{59}& 0   &  200 &  250 & 800 & \cdots & 2540\\

\VMSR{} &  0.8631   &  0.8568& 0.8560 & 0.8352 & \cdots & 0.6074\\

\Xhline{1pt}

\end{tabular*}
\end{table}

c) When the \VPbus{59} approaches to the critical active power point \VPbusmax{59}, a little step change of \VPbus{59} will lead to a small value of \VMSRmin{}. The feature b) and c) is available to conduct vulnerable node identification \cite{zhao2014research} as a new method. And when  \VPbus{59} is beyond \VPbusmax{59} (i.e. \Data {\VPbus{59}> 2555} MW), \VMSR{} is no longer steady.

\begin{table}[!h]  
\centering
\begin{tabular*}{8.8cm} { l  !{\color{black}\vrule width1pt} >{$}l<{$} >{$}l<{$} >{$}l<{$}  >{$}l<{$}  }  

\Xhline{1.5pt}

\VDelta {\VPbus{59}}& 50  &  50  & 40 & 15\\

\VPbus{59}&    200\!\rightarrow{}\!250 &  800\!\rightarrow{}\!850 &2500\!\rightarrow{}\!2540 &2540\!\rightarrow{}\!2555 \\

\VMSRmin{} &  0.6522   &  0.6122 & 0.4196 & unsteady\\

\VDelta{\VMSR{}} &  0.2046   &  0.2230 & 0.2085 & unsteady\\

\Xhline{1pt}

\end{tabular*}
\end{table}

\subsection{Case 3: Critical Power Point Estimation}

\begin{figure}[!h]
\centering
\subfloat[\Egam{\Text {Acc}}{0}, \Egam{\Text {Mul}}{0}]{
\begin{overpic}[width=0.4\textwidth
]{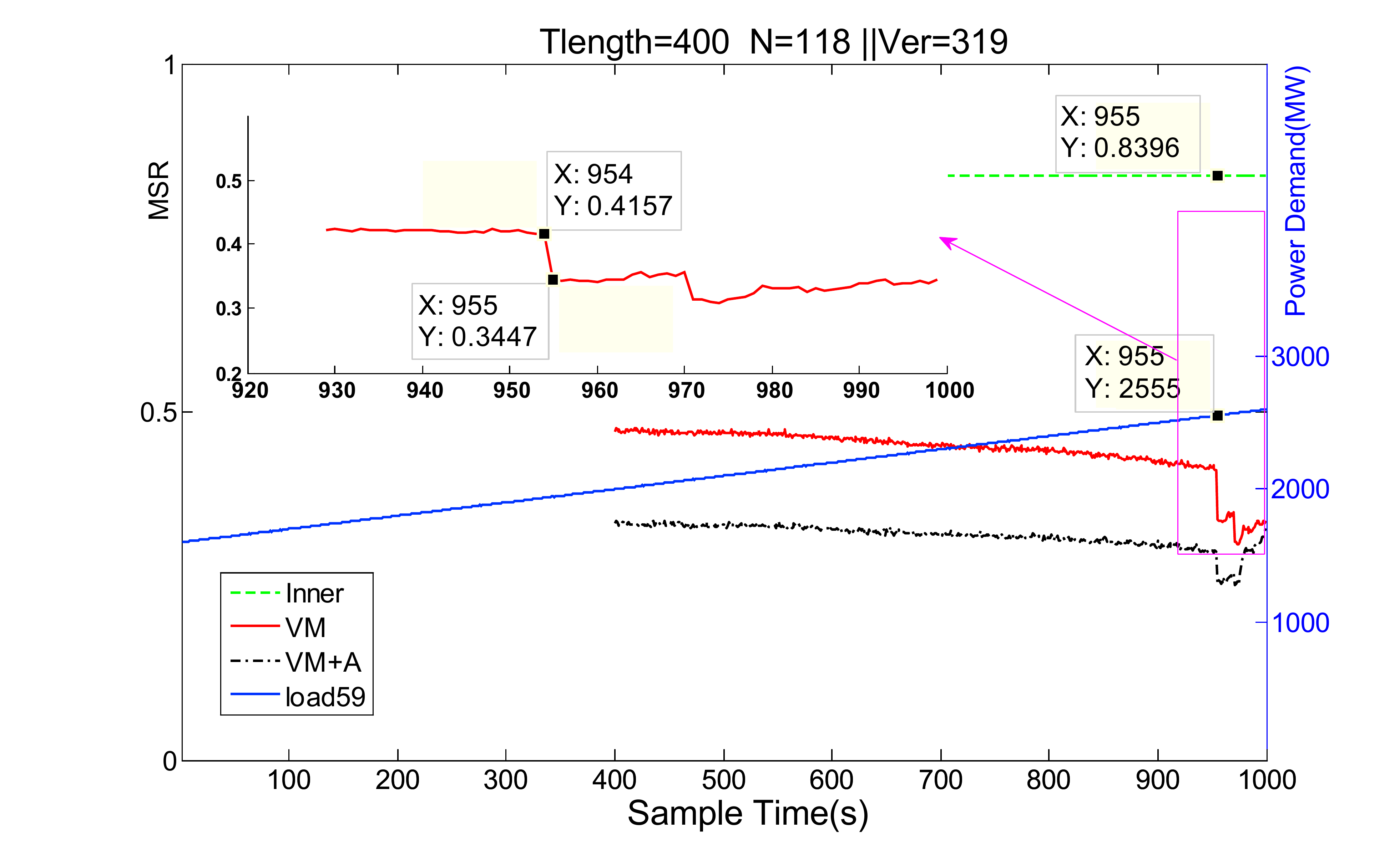}
    \setlength {\fboxsep}{1pt}
    \put(25,10) {\fcolorbox{red}{white}{\TI {\EPbusmax{59}{2555}{MW}}}  }
\end{overpic}
\label{fig:case3a}
}

\subfloat[\Egam{\Text {Acc}}{1}, \Egam{\Text {Mul}}{0.02}]{
\begin{overpic}[width=0.4\textwidth
]{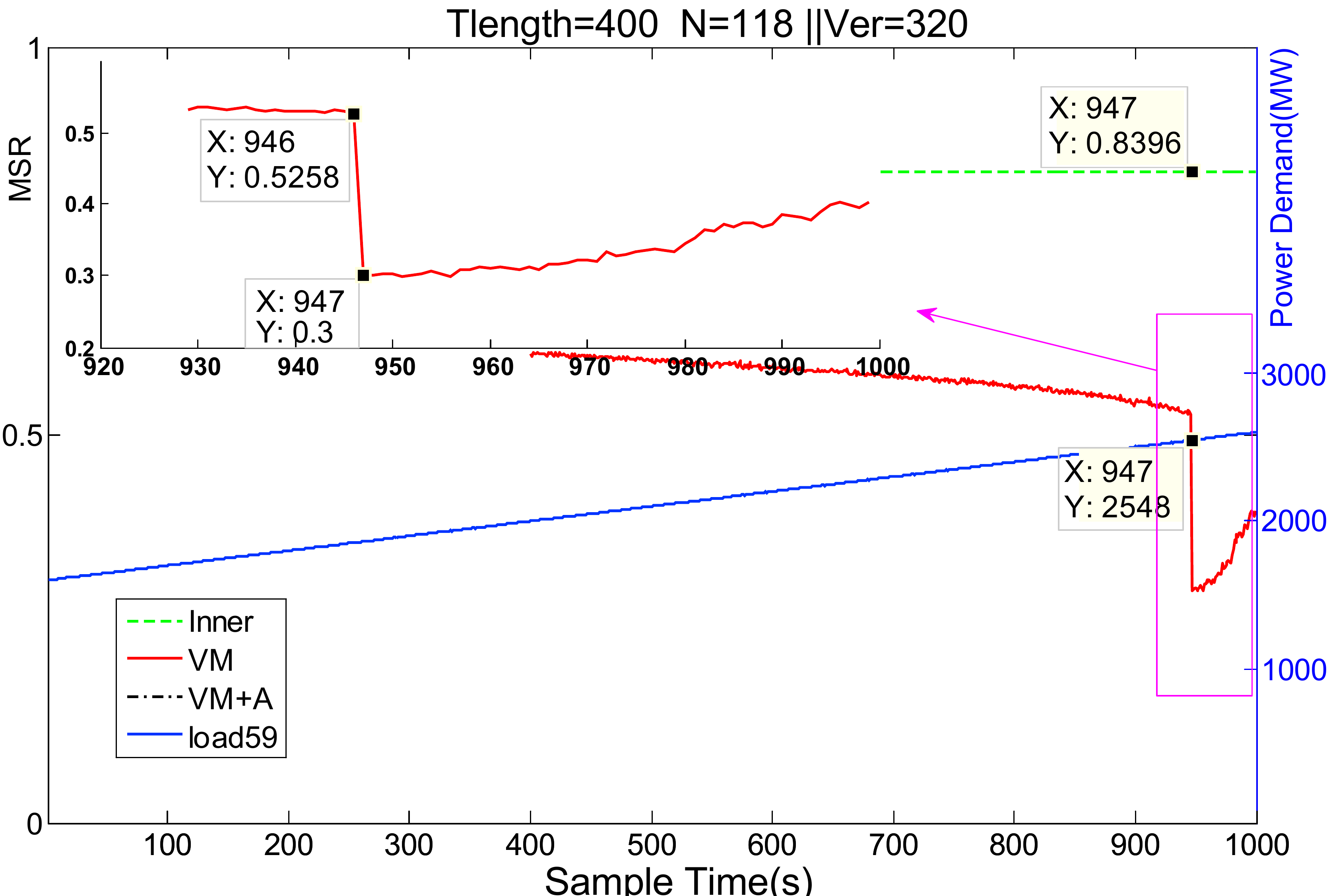}
    \setlength {\fboxsep}{1pt}
    \put(25,10) {\fcolorbox{red}{white}{\TI {\EPbusmax{59}{2548}{MW}}}  }
\end{overpic}
\label{fig:case3b}
}

\subfloat[\Egam{\Text {Acc}}{5}, \Egam{\Text {Mul}}{0.1}]{
\begin{overpic}[width=0.4\textwidth
]{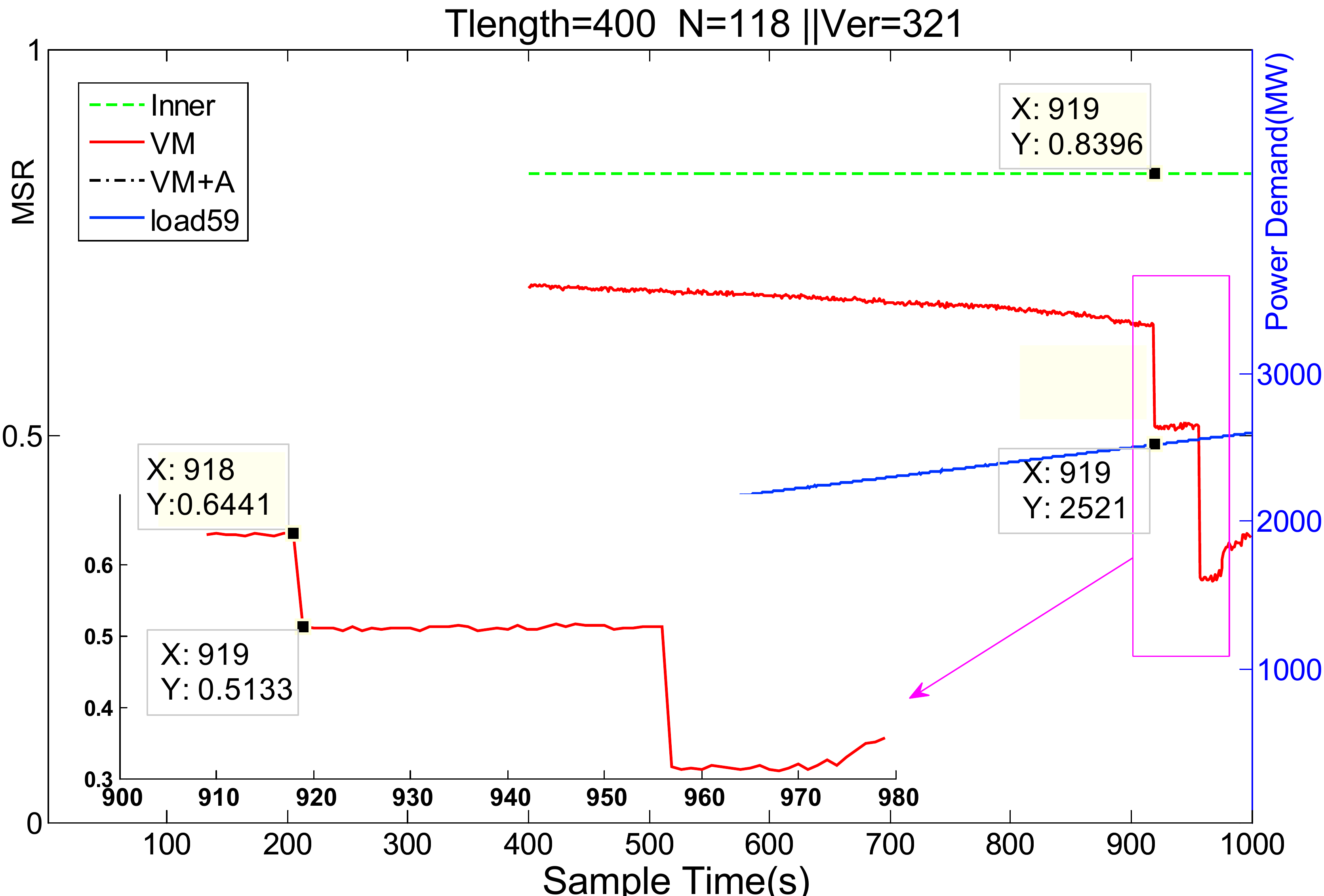}
    \setlength {\fboxsep}{0.1pt}
    \put(5,38) {\fcolorbox{red}{white}{\TI {\EPbusmax{59}{2526}{MW}}}  }
\end{overpic}
\label{fig:case3a}
}

\caption{Critical Power Point Estimation Taking Account of Grid Fluctuations}
\label{fig:case3}
\end{figure}

This case, based on the feature c) of the previous one, is designed as a new method to find the critical point \VPbusmax{n}, especially that the grid fluctuations is taking into consideration. The grid fluctuations are set by \Vgam{\Text {Acc}} and \Vgam{\Text {Mul}}
\FuncC{eq:gridfluctuation}{
\Equs {\Index {\Tdata y}{\Text{load\_}nt}}  {
\Index  y{\Text{load\_}nt}\Mul{}{}(1+\Muls{\Vgam{\Text {Mul}}}{\Vsx 1})
} +   \Muls{\Vgam{\Text {Acc}}}{\Vsx 2}
} 		
where \Vsx 1 and \Vsx 2 are random numbers from a standard Gaussian Distribution.
The result are shown in Fig. \ref{fig:case3}.

In this model, the increase of grid fluctuations means the decrease of signal-noise ratio, which will cause a raise of \VMSRmin{}. Meanwhile, it will also cause the decrease of \VPbusmax{n} for a certain node ($2555$ MW, $2548$ MW, $2521$ MW), which meet our common knowledge and experience.

\subsection{Case 4: Group-work Mode}
For the above cases, the anomaly can be detected by traditional model-based tools. These one validate that the designed data-driven architecture is also a solution to anomaly detection. In Case 4, however, we will design a case which can hardly be solved by traditional tools.

\begin{figure}[!h]
\centering
\subfloat[\VIndex A{\Text{Node}}=A1]{
\includegraphics[width=0.38\textwidth]{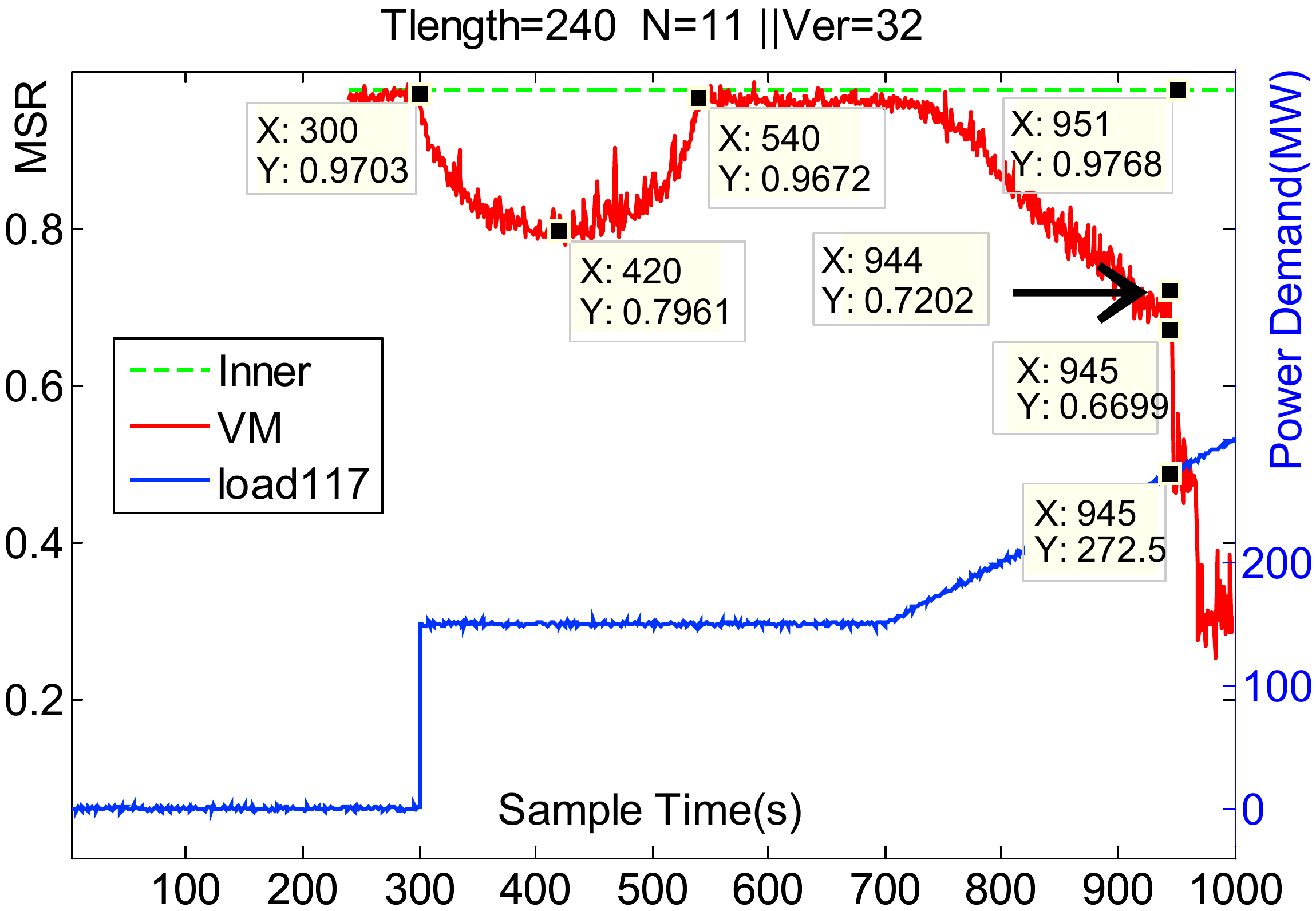}
\label{fig:case4c}
}

\subfloat[\VIndex A{\Text{Node}}=A2]{
\includegraphics[width=0.38\textwidth]{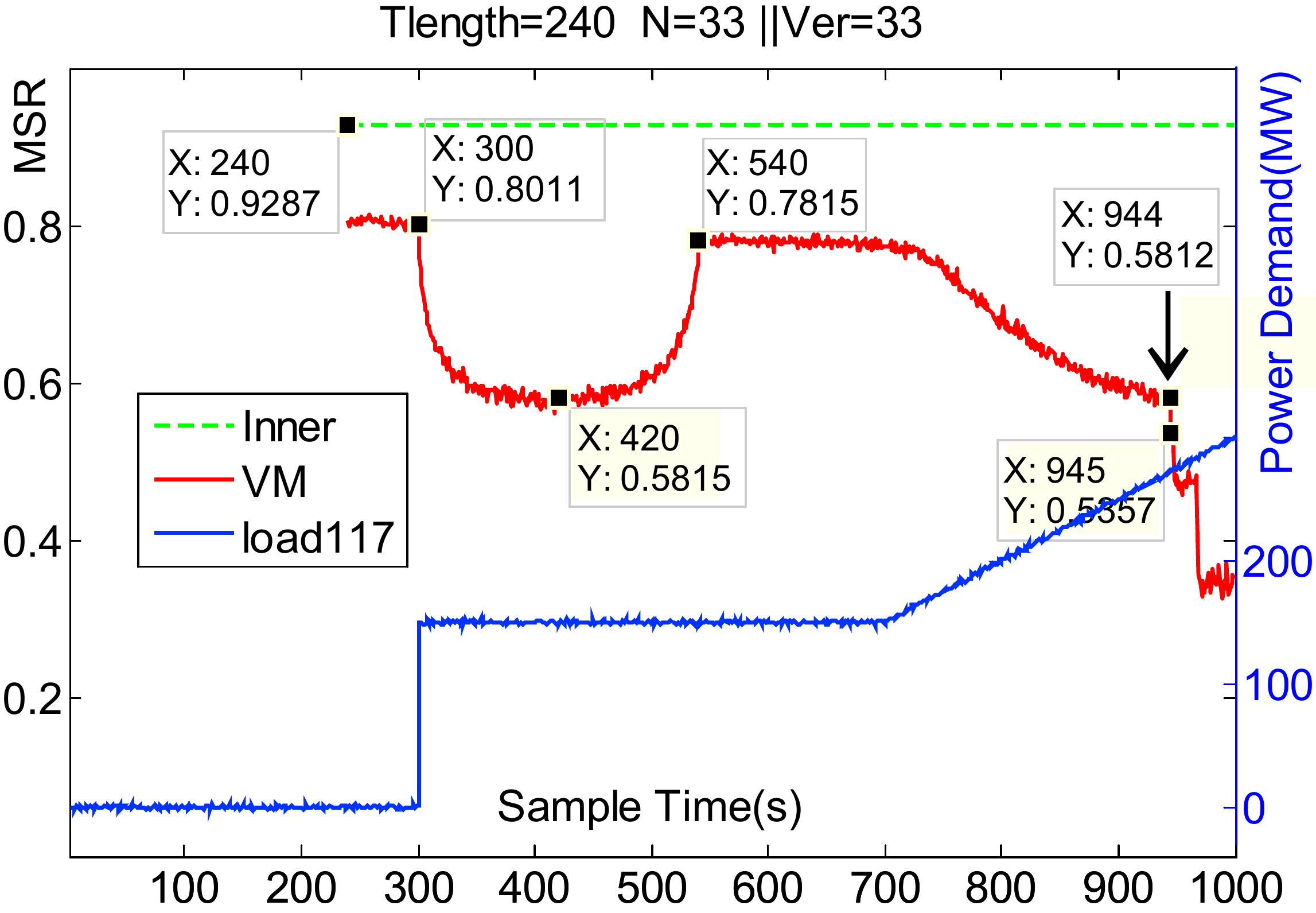}
\label{fig:case4d}
}
\caption{MSR Series of Single Region}
\label{fig:case4cd}
\end{figure}

\begin{figure}[H]
\centering
\begin{overpic}[scale=0.58
]{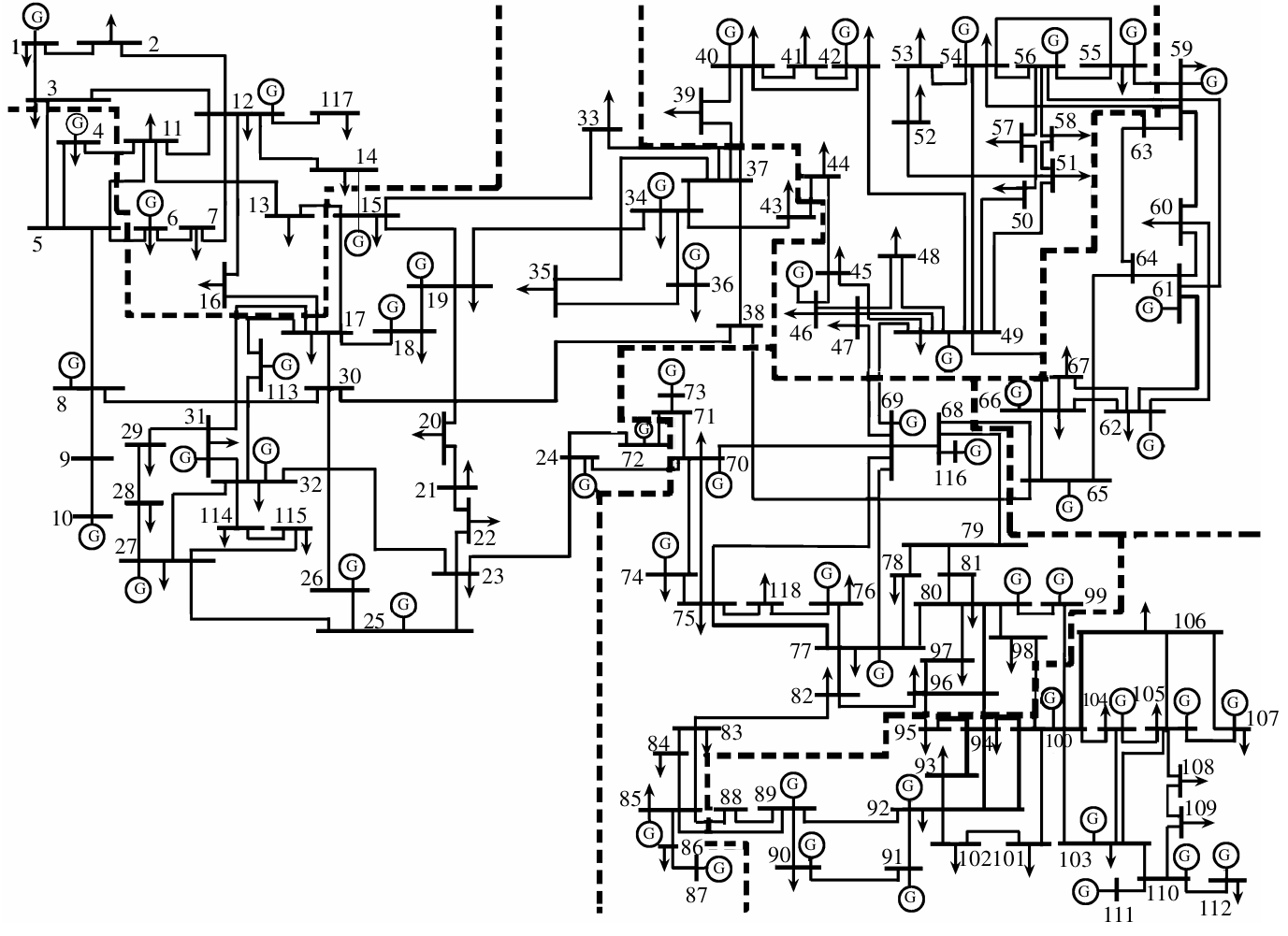}
    \setlength {\fboxsep}{1pt}
    \put(25,67) {\fcolorbox{red}{white}{\tiny \color{blue}{$117$}}} 
    \put(93,70) {\fcolorbox{red}{white}{\tiny \color{blue}{$59$}}} 
      \setlength {\fboxsep}{2pt}
      \put(32,70) {\fcolorbox{white}{ssBlue}{\tiny \color{black}{A1}}}
      \put(20,20) {\fcolorbox{white}{ssOrange}{\tiny \color{black}{A2}}}
      \put(41,68) {\fcolorbox{white}{ssOrange}{\tiny \color{black}{A2}}}
      \put(02,50) {\fcolorbox{white}{ssOrange}{\tiny \color{black}{A2}}}
      \put(50,70) {\fcolorbox{white}{sOrange}{\tiny \color{black}{A3}}}
      \put(92,34) {\fcolorbox{white}{Orange}{\tiny \color{black}{A4}}}
      \put(52,20) {\fcolorbox{white}{sBlue}{\tiny \color{black}{A5}}}
      \put(92,27) {\fcolorbox{white}{Blue}{\tiny \color{black}{A6}}}
      \put(75,02) {\fcolorbox{white}{Blue}{\tiny \color{black}{A6}}}

\end{overpic}
\caption{Partitioning network for IEEE 118-bus system. There are six partitions, i.e. A1, A2, A3, A4, A5, and A6.}
\label{fig:IEEE118network}
\end{figure}

\begin{figure}[H]
\centering
\includegraphics[width=0.4\textwidth]{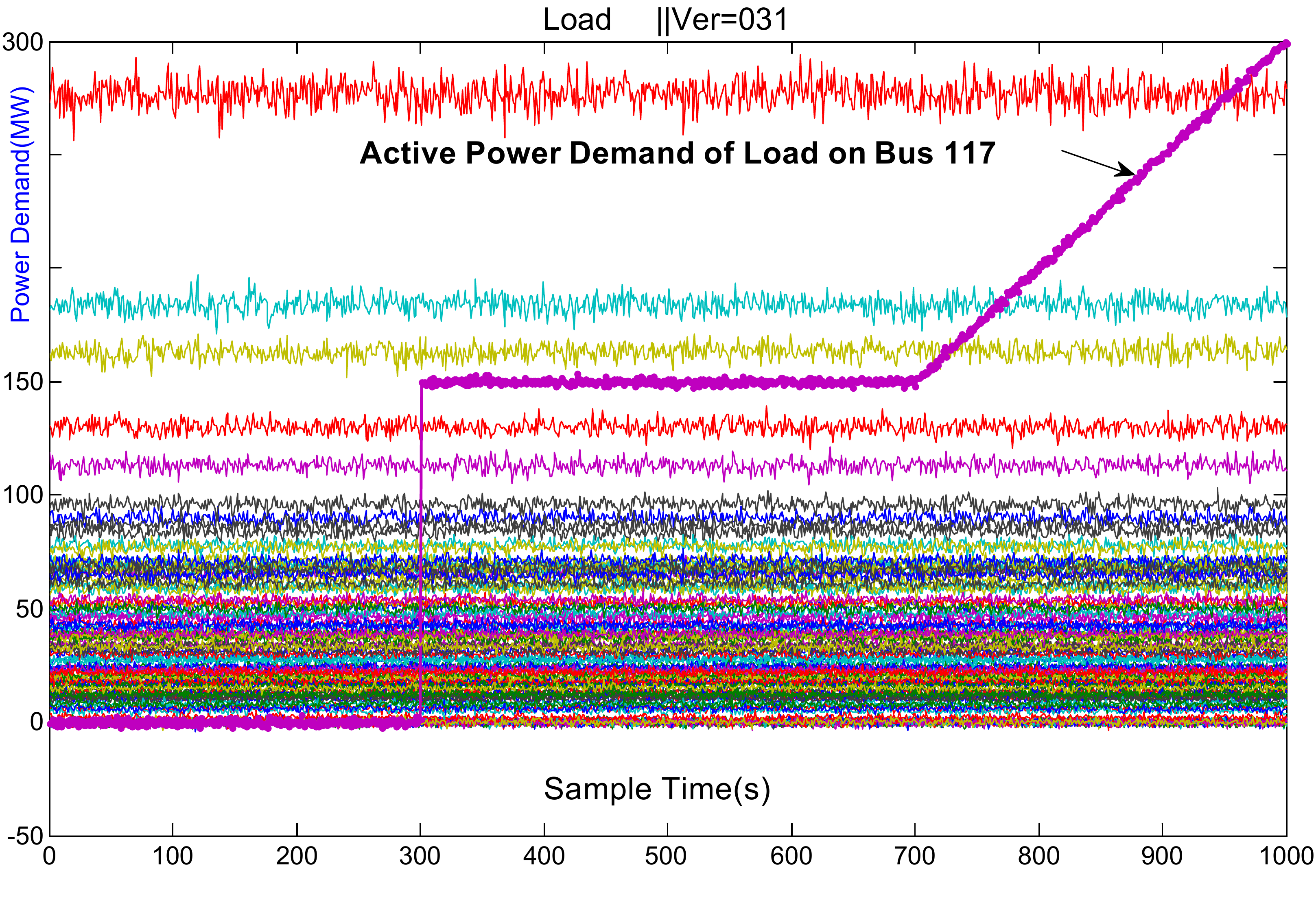}
\caption{Grid Load Change for Case 4: \Egam{\Text {Acc}}{1}, \Egam{\Text {Mul}}{0.02}}
\label{fig:case4a}

\includegraphics[width=0.48\textwidth]{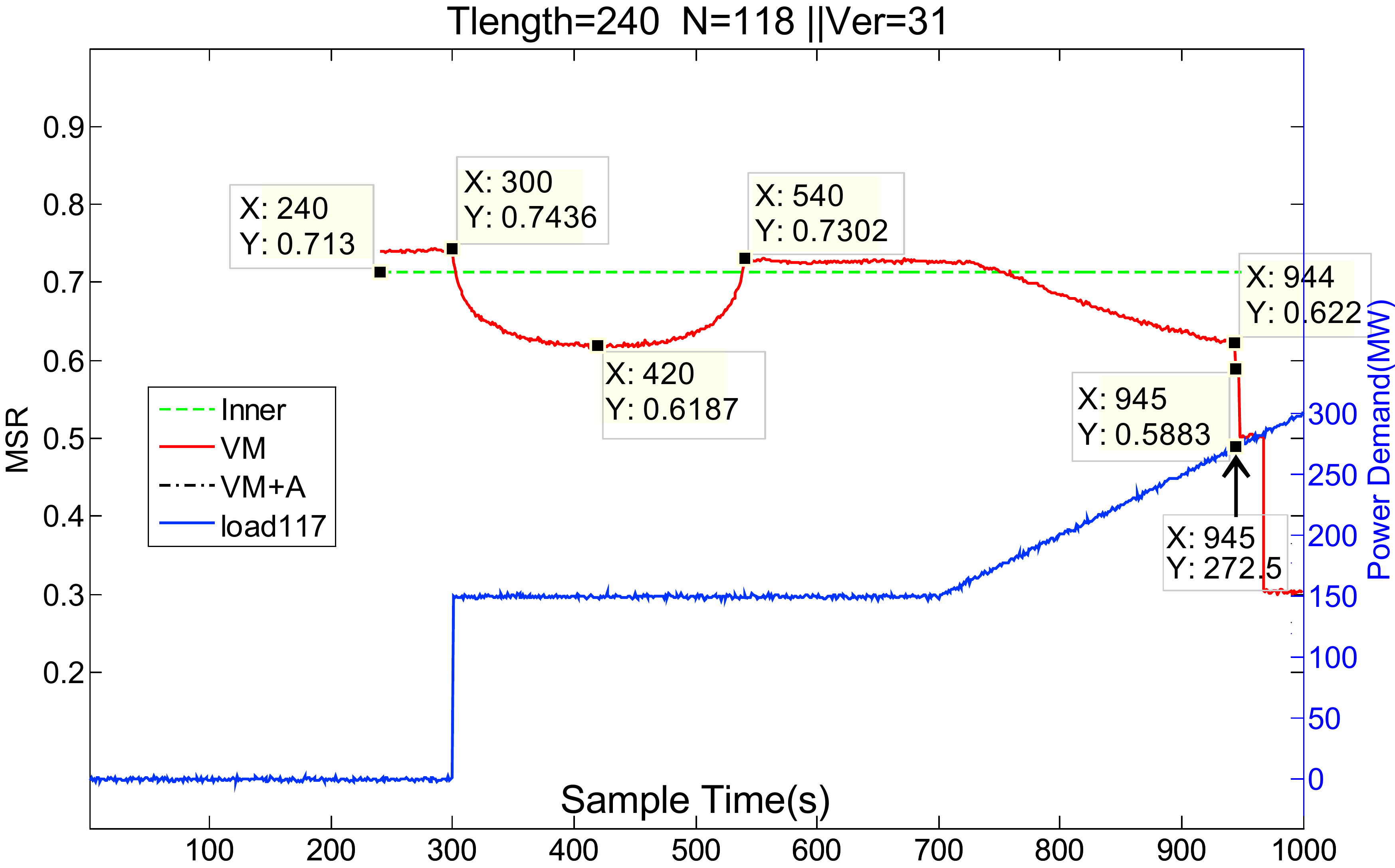}
\caption{Global MSR Series(\EPbusmax{117}{272.5}{MW})}
\label{fig:case4b}
\end{figure}

 This case is based on a zone-dividing system with 6 regions (A1 to A6) depicted in Fig. \ref{fig:IEEE118network}. A PQ node far from slack bus, i.e. bus-117 in A1, is chosen as signal source. It is much more vulnerable than PV nodes such as bus-59. With the same procedures of the former case studies, \VPbusmax{117} and  the \Cur {\VMSR{}}{t} curve for the overall system are depicted by Table \ref{tab:case4}, Fig. \ref{fig:case4a}, and Fig. \ref{fig:case4b}, respectively.
 Under group-work mode, each regional center, just like the global one, calculates MSR with its own data, such as Fig.  \ref{fig:case4c} for A1, and  \ref{fig:case4d} for A2.

\begin{figure}[H]
\centering
\subfloat[{Raw Data \VRV{}} (Orange for A3, and Blue for A5)]{
\includegraphics[width=0.48\textwidth]{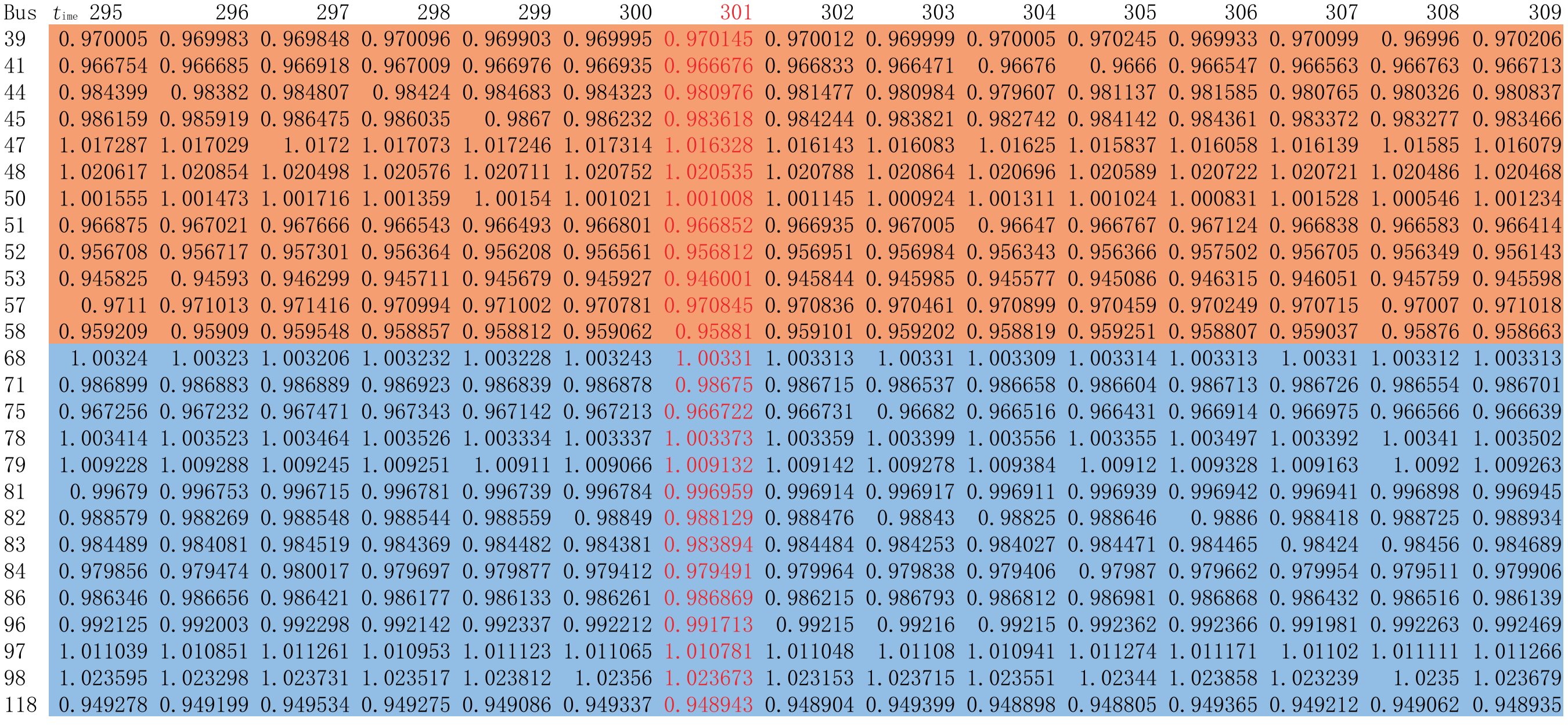}
\label{fig:case4o}
}

\subfloat[{Visualization of the above \VRV{} (in low dimension)}]{
\includegraphics[width=0.48\textwidth]{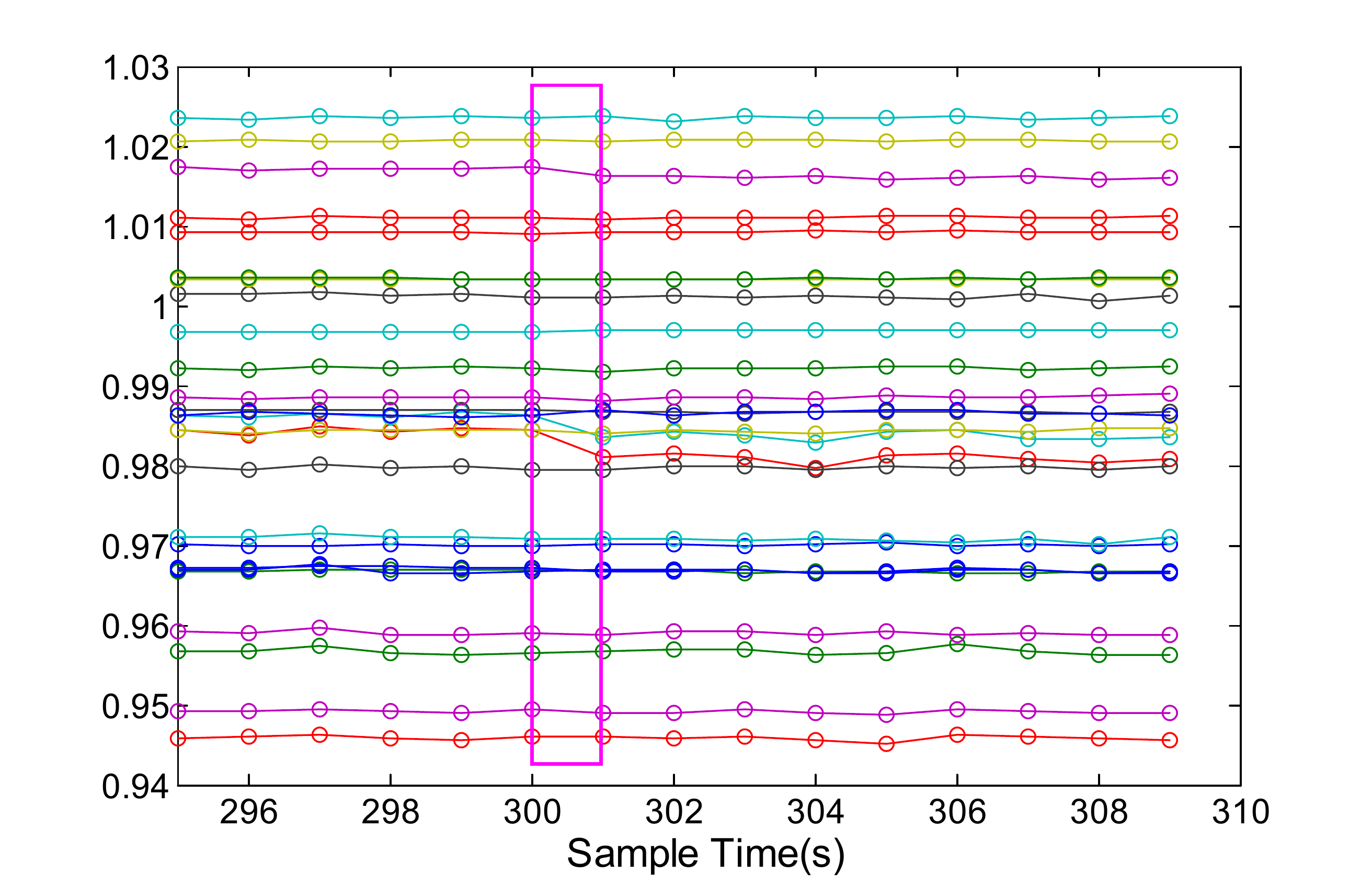}
\label{fig:case4p}
}

\caption{Raw Data \VRV{} and the Visualization around \EtS{300}{s}}
\label{fig:case4op}
\end{figure}

When the data split-window is not big, just as A1 which only has 11 nodes (i.e. \Equs{N}{11}), the signal is still able to detected, but the curve is not smooth. Thus, some regions are combined to smooth  the \Cur{\VMSR{}}{t} curve---A1\&A2 in Fig. \ref{fig:case4h}, A3\&A5 in \ref{fig:case4i}, and A4\&A6 in \ref{fig:case4j}.  In the appendices, the raw data of \VRV{} and their low-dimensional visualization for all PQ buses in A3\&A5 around time \EtS{301}{s}, when the step-change of \VPbus{117} happened, are shown as Fig. \ref{fig:case4o}, \ref{fig:case4p}.

\begin{figure*}[htbp]
\centering
\subfloat[{\VIndex A{\Text{Node}}=[A1,A2]}]{
\includegraphics[width=0.31\textwidth]{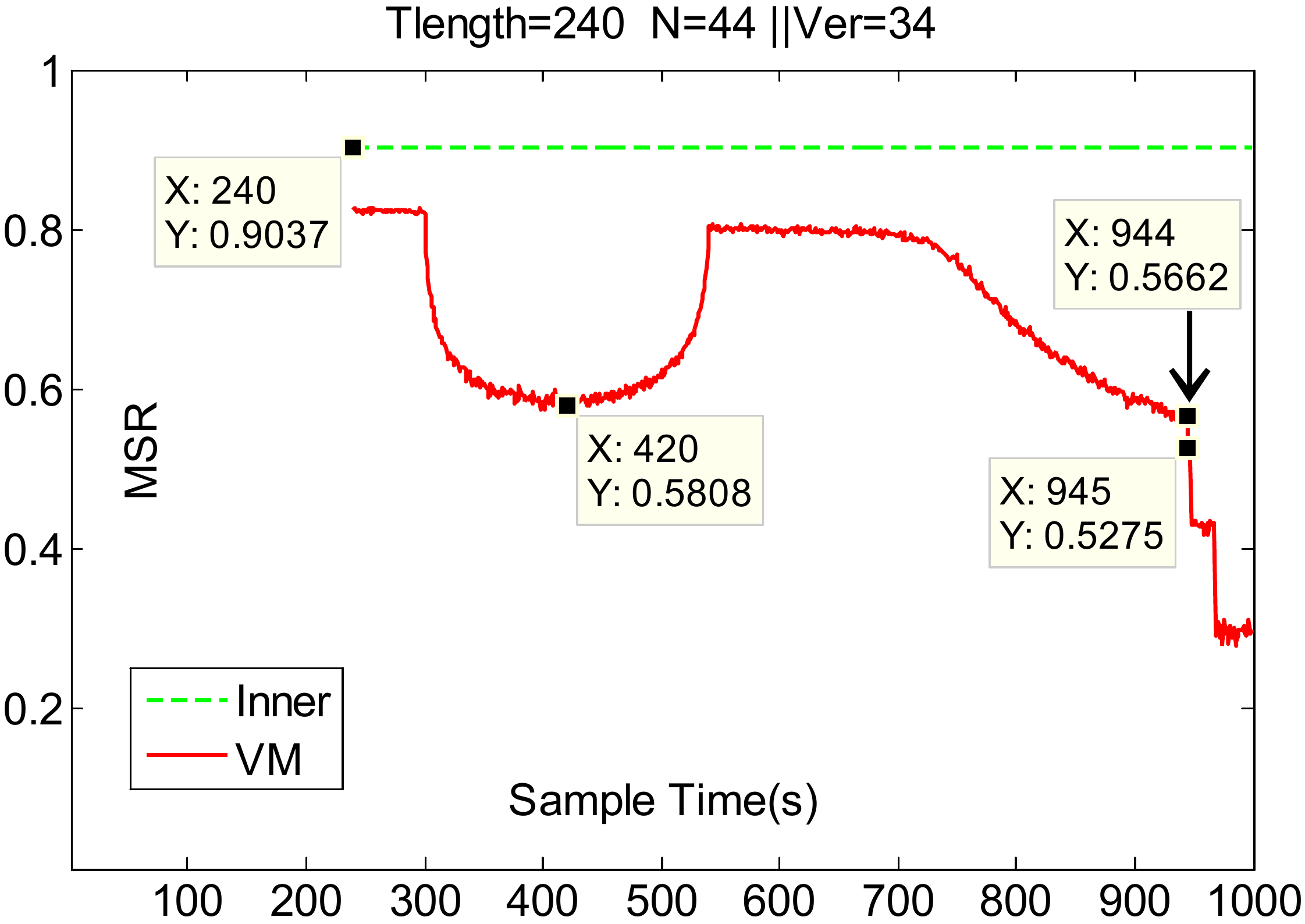}
\label{fig:case4h}
}
\subfloat[{\VIndex A{\Text{Node}}=[A3,A5]}]{
\includegraphics[width=0.31\textwidth]{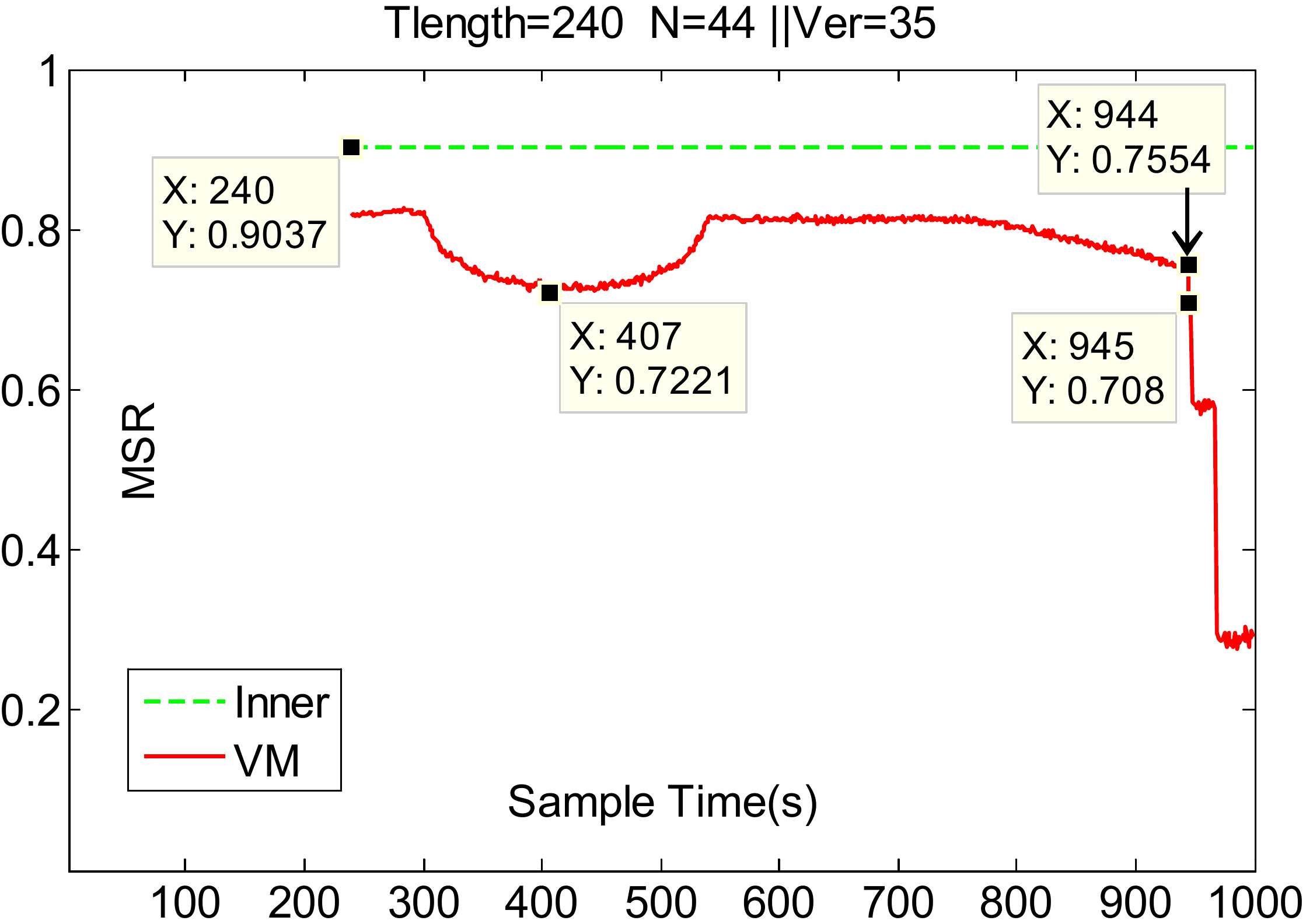}
\label{fig:case4i}
}
\subfloat[{\VIndex A{\Text{Node}}=[A4,A6]}]{
\includegraphics[width=0.31\textwidth]{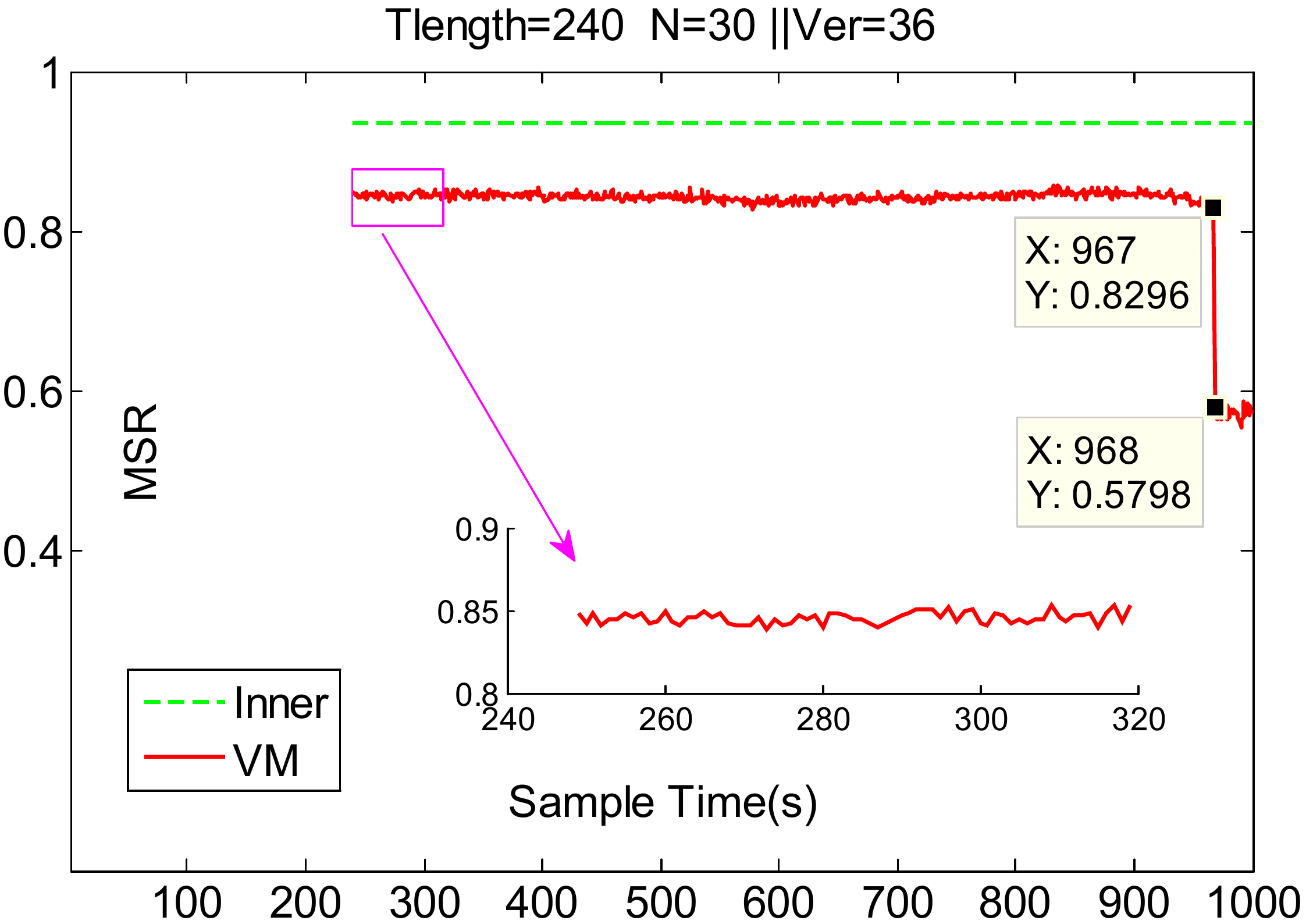}
\label{fig:case4j}
}
\caption{MSR Series of Union Regions}
\label{fig:case4hij}
\end{figure*}

Fig. \ref{fig:case4b} shows that \EPbusmax{117}{272.5}{MW} at \EtS{945}{s}. This point is observed by all the regional centers as shown in Fig. \ref{fig:case4cd} and  \ref{fig:case4hij}. Meanwhile, the signal occurring just at time \Et{301}{s} is also able to be detected in the system. According to \VMSR{} of Fig. \ref{fig:case4h}, \ref{fig:case4i} and \ref{fig:case4j}, it is found that to response to the signal, the distribution of \VDelta {\VMSR{}} in distributed regions is just like the contour line and the A1\&A2 is the mountaintop. As a result, we conjecture that the signal is generated at A1\&A2 rather than A3\&A5 or A4\&A6. The events set in Table  \ref{tab:case4} validates the conjecture.

Interchanging MSRs among distributed utilities, some useful analyses are able to be extracted. It is sensitive to anomaly detection, even with imperceptible different raw data just as shown in Fig. \ref{fig:case4op}---the raw voltage amplitude dataset \VOG{\VRV{}} of A3\&A5 changes too little to be utilized in low-dimension (e.g., Mean---one-dimensional statistic, Variance---two-dimensional statistic).
This case study also validates the architecture is compatible with block calculation only using regional small database; in this way, the architecture decouples the interconnect systems from statistical parameters perspective naturally, and is practical for real large-scale distributed systems.

\subsection{Case 5: Fault Detection for Active Distribution Network}
This case shows the application of the designed architecture in another field of power systems. Due to the integration and variation of renewable generators and energy storage units, faults and disturbances detection has become increasingly complicated in active distribution networks (ADN) \cite{huang2014impedance}. Table \ref{tab:case5} sets the events series, and Fig. \ref{fig:case5a} illustrates the fault mode.

Fig. \ref{fig:case5b} indicates that some signals are detected by the \Cur {\VMSR}{t} curve. Especially, it is conjectured that the most influent events are happening around \Et{3,000}{ms} and \Et{13,000}{ms}, which means that the three-phase and line-to-line short circuit have more influence than the single-phase ones. This case shows that the architecture is also compatible with other fields in power systems.

\Figf {fig:case5b}{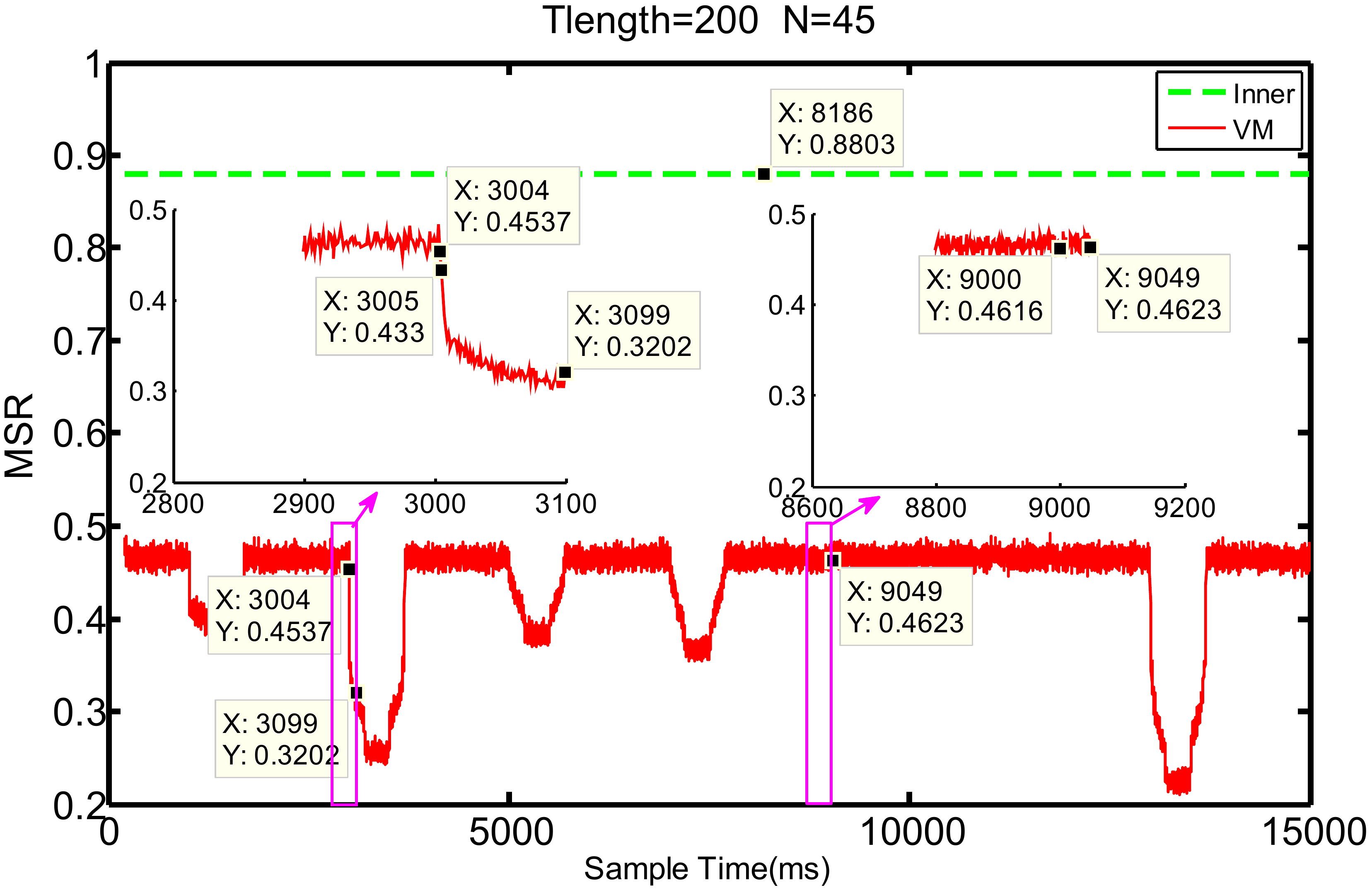}{MSR Series for Failure Events}{htbp}

\section{Conclusion}
This paper aims to apply big data technology into smart grids. It proposes an architecture with two independent procedures, as a data-driven solution, to conduct anomaly detections. In addition, moving split-window technology was introduced for real-time analysis, and a new statistic mean spectral radius was proposed to indicate the data correlations, as well as to clarify  the parameter interchanged among the utilities.

The algorithm of this architecture is based on random matrix theory (RMT); it is a fixed objective procedure, which is easy in logic and fast in speed. The non-asymptotic framework of RMT enables us to conduct high-dimensional analysis for real systems, even with relatively moderate datasets. It also provides us a natural way to decouple the interconnected systems from data perspective. The group-work model of utilities in the systems, meanwhile, makes some data-driven functions possible, such as distributed calculation and comparative analysis.

Big data technology does not conflict with classical analysis or pretreatment. Instead, combination between block calculation and traditional zone-dividing structure realizes comparative analysis---it is sensitive way to detect the event and locate the source in the grid network, even with imperceptible different measured/simulated data. Besides, the sparse representation of  a random vector is carried out with the random matrix theory~\cite{xu2014sparse, pfander2008identification}. An incomplete matrix data representation of data may be considered: 1) random band matrices \cite{jana2014fluctuations}; 2) low rank matrix recovery \cite{qiu2013bookcogsen}. All these topics are considered through the unified framework of random matrix theory \cite{qiu2013bookcogsen}.

It is a long way and big topic to apply big data in power systems, especially for a specific field in real systems. Some designed methods are rough in this paper, e.g., \textit{to estimate critical power point} in \textit{case 3}, \textit{to detect fault in power systems} in \textit{case 5}, as well as some descriptions, e.g., \textit{group-work mode} mentioned in \textit{section III}, and \textit{new method for vulnerable node identification} mentioned in \textit{case 2}.  But these special methods/applications all have one thing in common: they are all based on the proposed architecture, and driven only by data of voltage amplitude, which are the most basic in the grids.
With work ongoing, including 1) designing 3D Power-map by combining high-dimensional analysis and visualization \cite{he20153d}, and 2) conducting event/fault detection for real systems \cite{he2015unsup,he2015fault}, we can say with confidence that the designed architecture is practical for real world; and the keys are the RMT with non-asymptotic framework, high-dimensional algorithm, block calculation, and augmented matrix. This paper, as an initial one, just presents the universal architecture; for special fields, it aims to raises many open questions than actually answers ones.
One wonders if this new direction will be far-reaching in years to come toward the age of big data.

\appendices

\section{The Grid Network Structures of G1, G2 and G3}

\Figffb {fig:G3s}{G1}{G2}{G3}{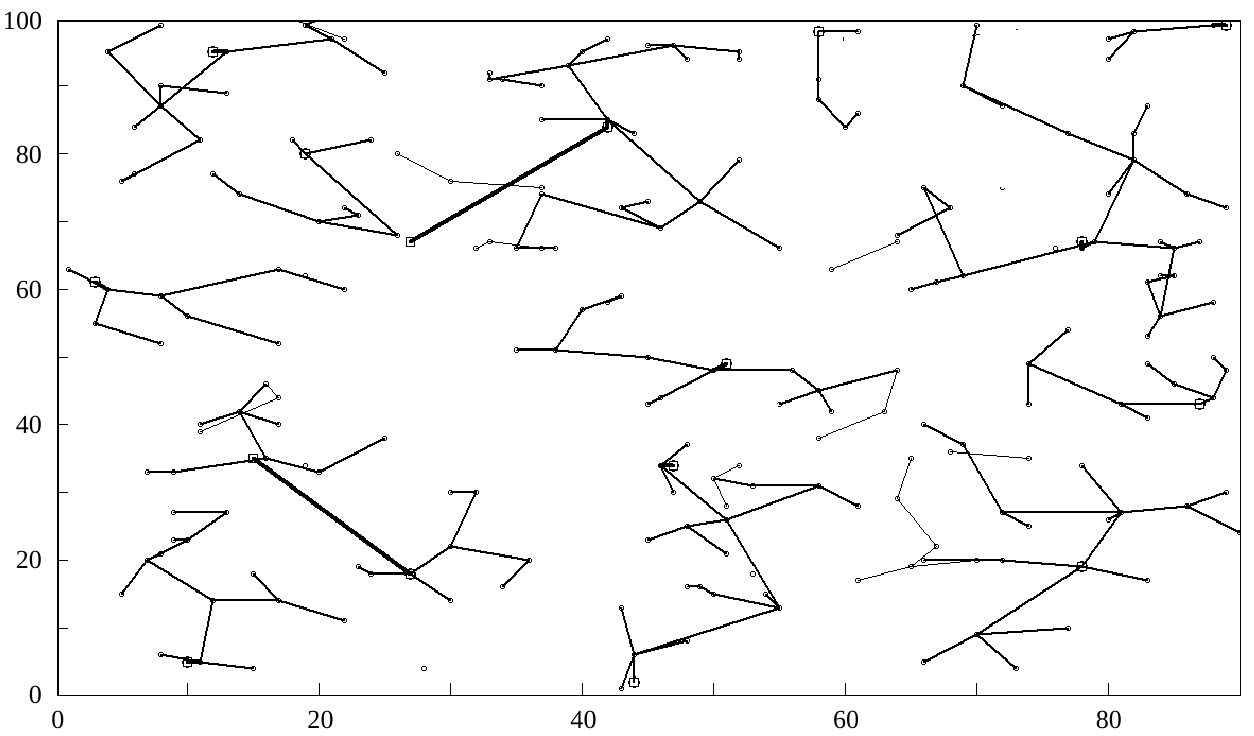}{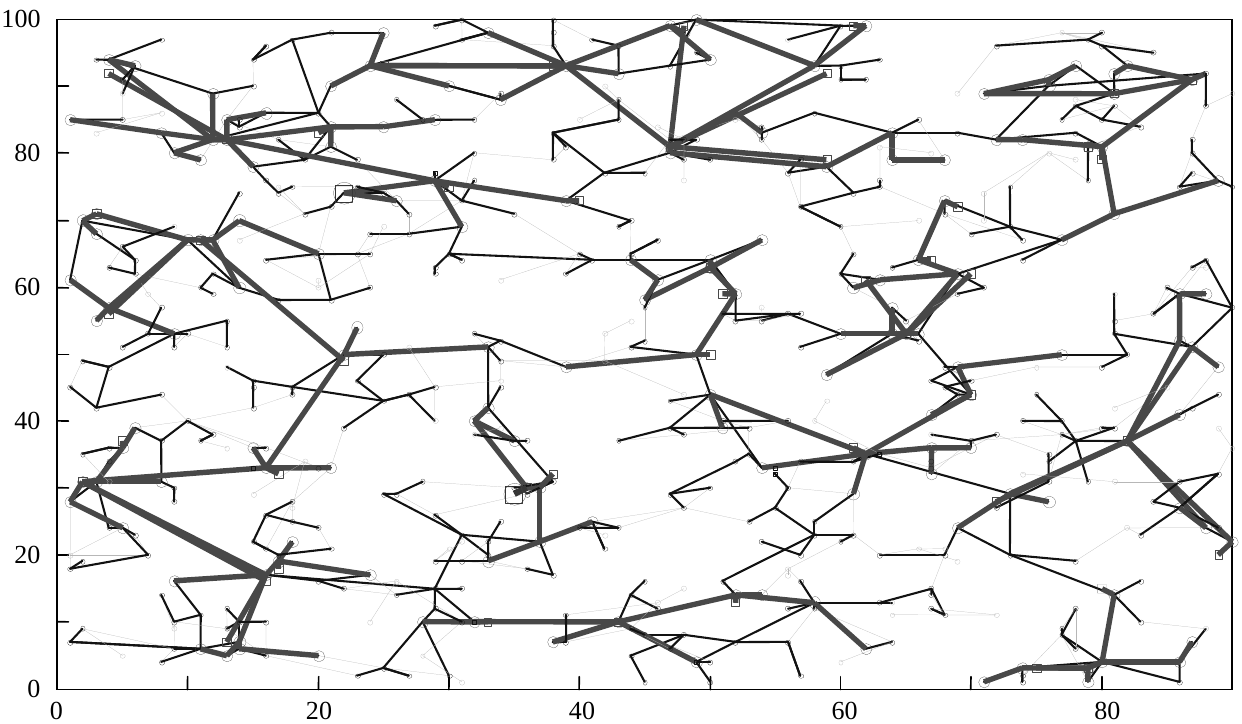}{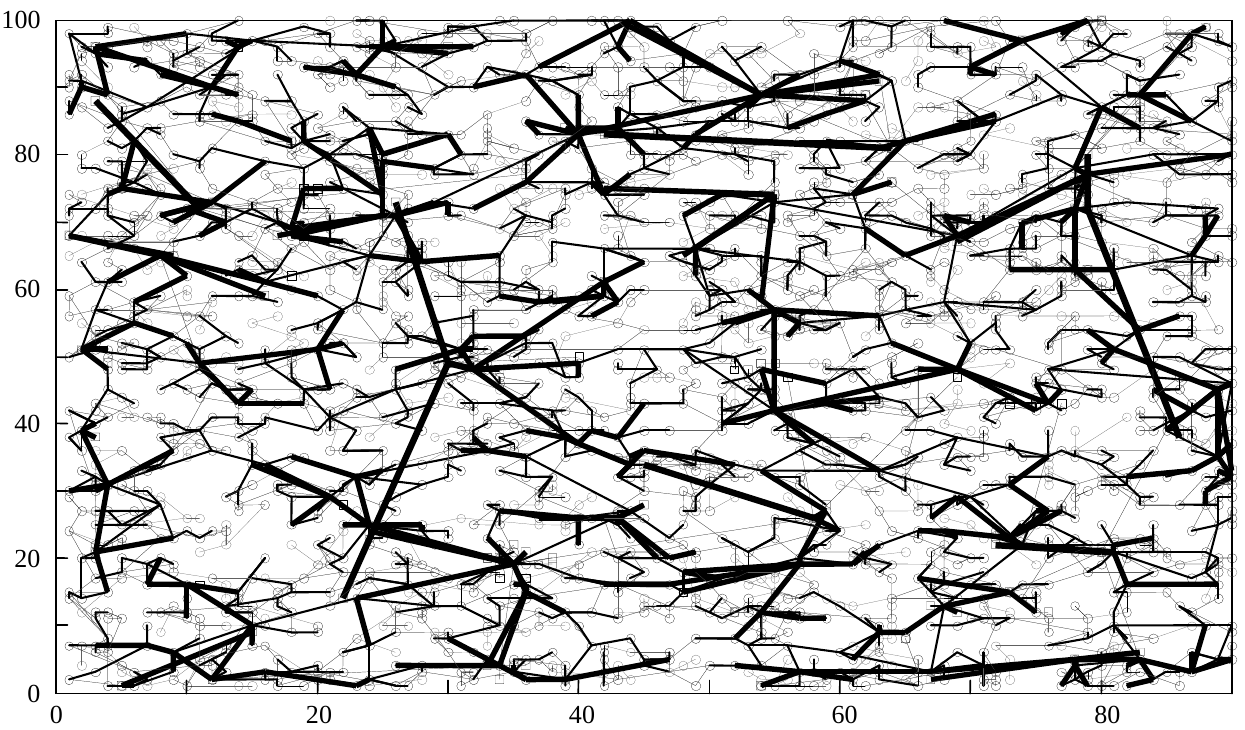}
{Simulation of network structures. The above twos are G1---Small-scale isolated grids, and G2---Large-scale interconnected power grids; and the below one is G3---Smart grids, which have complex network structures without clear-cut partitioning.}
{H}   
%

\section{Event Series for Five Case Studies}

\begin{table}[H]  
\caption {Series of Events  for Case 1}
\label{tab:case1es}
\centering

\begin{minipage}[!t]{0.48\textwidth}
\centering

\begin{tabularx}{\textwidth} { l!{\color{black}\vrule width1pt}    >{$}l<{$}    >{$}l<{$}   >{$}l<{$}   }  
\toprule[1.5pt]
\hline

\Vt{} & [001\!:\!550] & [551\!:\!1100] & [1101\!:\!1650]\\
\VPbus{59} & 0 & 200  & 0 \\

\hline
\Vt{}  & [1651\!:\!2200] & [2201\!:\!2500] &     \\
\VPbus{59} &  1500 & 0 &  \\

\toprule[1pt]

\end{tabularx}
\raggedright
\scriptsize {*\Vt{}: (s), \VPbus{59}: (MW)  }
\end{minipage}

\end{table}

\begin{table}[H]  

\caption {Series of Events for Case 2}
\label{tab:case2}
\centering

\begin{minipage}[!t]{0.48\textwidth}
\centering

\begin{tabularx}{\textwidth} { l!{\color{black}\vrule width1pt}    >{$}l<{$}    >{$}l<{$}   >{$}l<{$}   }  
\toprule[1.5pt]
\hline

\Vt{} & [001\!:\!300] & [301\!:\!600] & [601\!:\!900]\\
\VPbus{59} & 0 & 200  & 250 \\

\hline
\Vt{}  & [901\!:\!1200] & [1201\!:\!1500] & [1501\!:\!1800]    \\
\VPbus{59} & 800 & 850 & 2000 \\

\hline
\Vt{}  & [1801\!:\!2100] & [2101\!:\!2400] & [2401\!:\!2500]    \\
\VPbus{59} &  2540 & 2555 & 3500 \\

\toprule[1pt]

\end{tabularx}
\raggedright
\scriptsize {*\Vt{}: (s), \VPbus{59}: (MW)  }
\end{minipage}

\caption {Series of Events for Case 4}
\label{tab:case4}
\centering

\begin{minipage}[!t]{0.48\textwidth}
\centering

\begin{tabularx}{\textwidth} { l !{\color{black}\vrule width1pt}  >{$}l<{$} >{$}l<{$}  >{$}l<{$} }
\toprule[1.5pt]
\hline

\Vt{} & [001\!:\!300] & [301\!:\!700] & [701\!:\!1000]\\
\VPbus{117} & 0 & 150  & {t/2-200} \\
\toprule[1pt]
\end{tabularx}
\raggedright
\scriptsize {*\Vt{}: (s), \VPbus{117}: (MW)  }
\end{minipage}

\caption {Series of Events for Case 5}
\label{tab:case5}
\centering

\begin{minipage}[!t]{0.48\textwidth}
\centering

\begin{tabularx}{\textwidth} { l !{\color{black}\vrule width1pt}  >{$}l<{$} >{$}l<{$}  >{$}l<{$} }
\toprule[1.5pt]
\hline

\Vt{} & [1000\!:\!1500] & [3000\!:\!3500] & [5000\!:\!5500]\\
\scshape {Event} & \Text A \rightarrow{} \Text G & \Text {AB} \rightarrow{} \Text G  & \Text B \rightarrow{} \Text G \\
\hline
\Vt{} & [7000\!:\!7500] & [9000\!:\!11000] & [13000\!:\!13500]\\
\scshape {Event} & \Text C \rightarrow{} \Text G & \Text {BCS OPEN}  & \Text {ABC}\rightarrow{} \Text G \\
\toprule[1pt]
\end{tabularx}
\raggedright
\scriptsize {*\Vt{}: (ms)}
\end{minipage}
\end{table}

\Figf {fig:case5a}{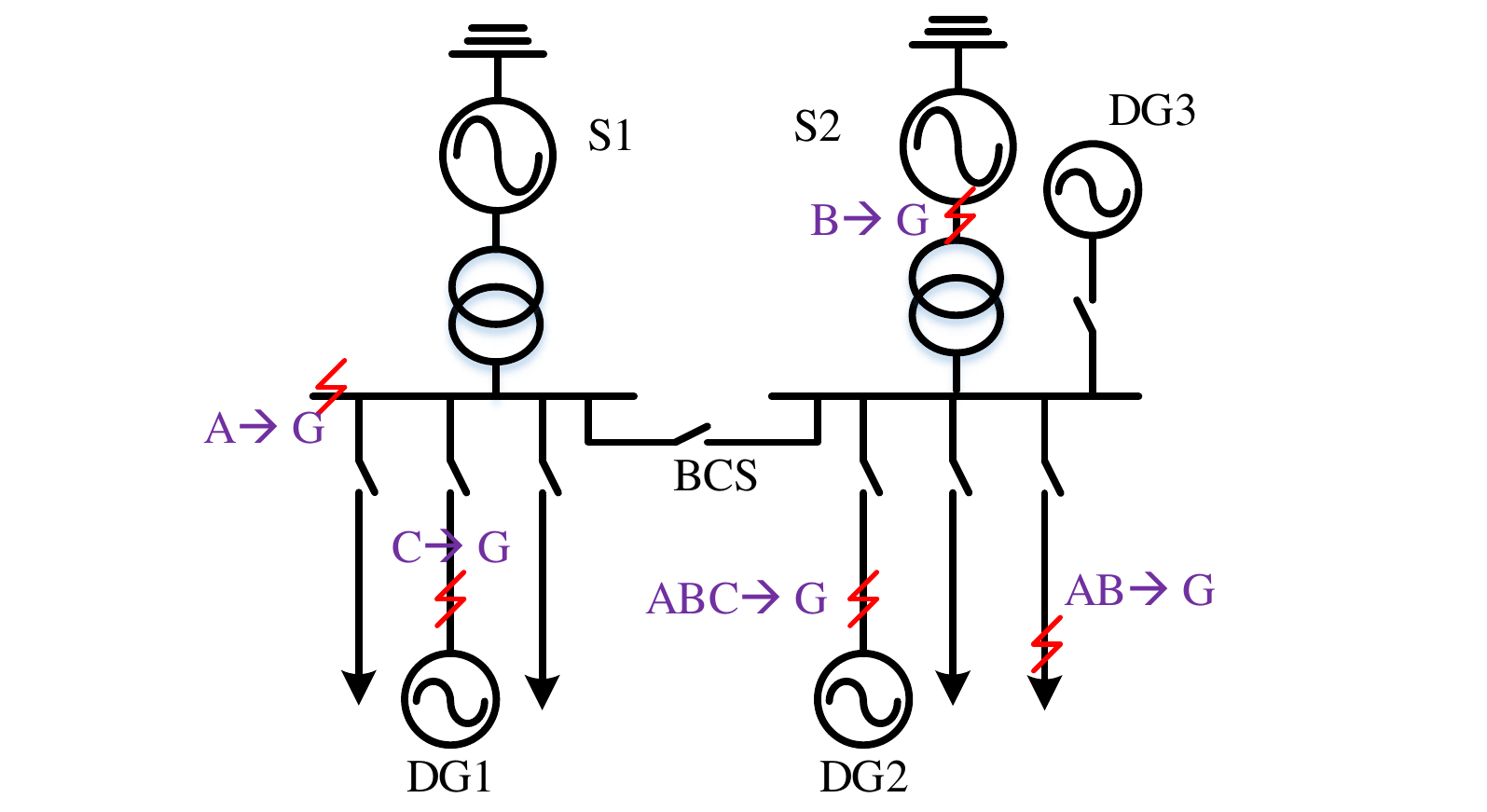}{Fault Model for Case 5}{H}   

\section{}

\bibliographystyle{IEEEtran}
\bibliography{helx}

\begin{thebibliography}{10}
\providecommand{\url}[1]{#1}
\csname url@samestyle\endcsname
\providecommand{\newblock}{\relax}
\providecommand{\bibinfo}[2]{#2}
\providecommand{\BIBentrySTDinterwordspacing}{\spaceskip=0pt\relax}
\providecommand{\BIBentryALTinterwordstretchfactor}{4}
\providecommand{\BIBentryALTinterwordspacing}{\spaceskip=\fontdimen2\font plus
\BIBentryALTinterwordstretchfactor\fontdimen3\font minus
  \fontdimen4\font\relax}
\providecommand{\BIBforeignlanguage}[2]{{%
\expandafter\ifx\csname l@#1\endcsname\relax
\typeout{** WARNING: IEEEtran.bst: No hyphenation pattern has been}%
\typeout{** loaded for the language `#1'. Using the pattern for}%
\typeout{** the default language instead.}%
\else
\language=\csname l@#1\endcsname
\fi
#2}}
\providecommand{\BIBdecl}{\relax}
\BIBdecl

\bibitem{nature2008bigd}
Nature, ``Big data (specials),'' Sep 2008,
  \mbox{\url{http://www.nature.com/news/specials/bigdata/index.html}}.

\bibitem{science2011bigd}
Science, ``Special online collection: Dealing with data,'' Feb 2011,
  \mbox{\url{ http://www.sciencemag.org/site/special/data/}}.

\bibitem{IBM2014fourv}
IBM, ``The four v¡¯s of big data.'' [EB/OL],
  {\url{http://www.ibmbigdatahub.com/infographic/four-vs-big-data}}.

\bibitem{qiu2015smart}
R.~Qiu and P.~Antonik, \emph{Smart Grid and Big Data}.\hskip 1em plus 0.5em
  minus 0.4em\relax John Wiley and Sons, 2014.

\bibitem{IBM2009Manag}
IBM, ``Managing big data for smart grids and smart meters,'' May 2009.

\bibitem{kezunovic2013role}
M.~Kezunovic, L.~Xie, and S.~Grijalva, ``The role of big data in improving
  power system operation and protection,'' in \emph{Bulk Power System Dynamics
  and Control-IX Optimization, Security and Control of the Emerging Power Grid
  (IREP), 2013 IREP Symposium}.\hskip 1em plus 0.5em minus 0.4em\relax IEEE,
  2013, pp. 1--9.

\bibitem{brody1981random}
T.~A. Brody, J.~Flores, J.~B. French, P.~Mello, A.~Pandey, and S.~S. Wong,
  ``Random-matrix physics: spectrum and strength fluctuations,'' \emph{Reviews
  of Modern Physics}, vol.~53, no.~3, p. 385, 1981.

\bibitem{laloux2000random}
L.~Laloux, P.~Cizeau, M.~Potters, and J.-P. Bouchaud, ``Random matrix theory
  and financial correlations,'' \emph{International Journal of Theoretical and
  Applied Finance}, vol.~3, no.~03, pp. 391--397, 2000.

\bibitem{chen2012business}
H.~Chen, R.~H. Chiang, and V.~C. Storey, ``Business intelligence and analytics:
  From big data to big impact.'' \emph{MIS quarterly}, vol.~36, no.~4, pp.
  1165--1188, 2012.

\bibitem{howe2008big}
D.~Howe, M.~Costanzo, P.~Fey, T.~Gojobori, L.~Hannick, W.~Hide, D.~P. Hill,
  R.~Kania, M.~Schaeffer, S.~St~Pierre \emph{et~al.}, ``Big data: The future of
  biocuration,'' \emph{Nature}, vol. 455, no. 7209, pp. 47--50, 2008.

\bibitem{qiu2013bookcogsen}
R.~Qiu and M.~Wicks, \emph{Cognitive Networked Sensing and Big Data}.\hskip 1em
  plus 0.5em minus 0.4em\relax Springer, 2013.

\bibitem{qiu2014Intial70N}
C.~Zhang and R.~C. Qiu, ``Data modeling with large random matrices in a
  cognitive radio network testbed: Initial experimental demonstrations with 70
  nodes,'' \emph{arXiv preprint arXiv:1404.3788}, 2014.

\bibitem{qiu2014MIMO}
\BIBentryALTinterwordspacing
X.~Li, F.~Lin, and R.~C. Qiu, ``Modeling massive amount of experimental data
  with large random matrices in a real-time {UWB-MIMO} system,'' \emph{CoRR},
  vol. abs/1404.4078, 2014. [Online]. Available:
  \url{http://arxiv.org/abs/1404.4078}
\BIBentrySTDinterwordspacing

\bibitem{phadke2008wide}
A.~Phadke and R.~M. de~Moraes, ``The wide world of wide-area measurement,''
  \emph{Power and Energy Magazine, IEEE}, vol.~6, no.~5, pp. 52--65, 2008.

\bibitem{terzija2011wide}
V.~Terzija, G.~Valverde, D.~Cai, P.~Regulski, V.~Madani, J.~Fitch, S.~Skok,
  M.~M. Begovic, and A.~Phadke, ``Wide-area monitoring, protection, and control
  of future electric power networks,'' \emph{Proceedings of the IEEE}, vol.~99,
  no.~1, pp. 80--93, 2011.

\bibitem{xie2012distributed}
L.~Xie, Y.~Chen, and H.~Liao, ``Distributed online monitoring of quasi-static
  voltage collapse in multi-area power systems,'' \emph{Power Systems, IEEE
  Transactions on}, vol.~27, no.~4, pp. 2271--2279, 2012.

\bibitem{kanao2005power}
N.~Kanao, M.~Yamashita, H.~Yanagida, M.~Mizukami, Y.~Hayashi, and J.~Matsuki,
  ``Power system harmonic analysis using state-estimation method for japanese
  field data,'' \emph{Power Delivery, IEEE Transactions on}, vol.~20, no.~2,
  pp. 970--977, 2005.

\bibitem{alahakoon2013advanced}
D.~Alahakoon and X.~Yu, ``Advanced analytics for harnessing the power of smart
  meter big data,'' in \emph{Intelligent Energy Systems (IWIES), 2013 IEEE
  International Workshop on}.\hskip 1em plus 0.5em minus 0.4em\relax IEEE,
  2013, pp. 40--45.

\bibitem{xu2013power}
W.~Xu and J.~Yong, ``Power disturbance data analytics--new application of power
  quality monitoring data,'' \emph{Proceedings of the CSEE}, vol.~19, p. 013,
  2013.

\bibitem{qiu2012bookcogpp}
R.~C. Qiu, Z.~Hu, H.~Li, and M.~C. Wicks, \emph{Cognitive radio communication
  and networking: Principles and practice}.\hskip 1em plus 0.5em minus
  0.4em\relax John Wiley \& Sons, 2012.

\bibitem{qiu2014foundation}
R.~C. Qiu, ``The foundation of big data: Experiments, formulation, and
  applications,'' \emph{arXiv preprint arXiv:1412.6570}, 2014.

\bibitem{marvcenko1967distribution}
V.~A. Mar{\v{c}}enko and L.~A. Pastur, ``Distribution of eigenvalues for some
  sets of random matrices,'' \emph{Sbornik: Mathematics}, vol.~1, no.~4, pp.
  457--483, 1967.

\bibitem{pan2011universality}
G.~Pan, Q.~Shao, and W.~Zhou, ``Universality of sample covariance matrices: Clt
  of the smoothed empirical spectral distribution,'' \emph{arXiv preprint
  arXiv:1111.5420}, 2011.

\bibitem{guionnet2009single}
A.~Guionnet, M.~Krishnapur, and O.~Zeitouni, ``The single ring theorem,''
  \emph{arXiv preprint arXiv:0909.2214}, 2009.

\bibitem{benaych2013outliers}
\BIBentryALTinterwordspacing
F.~Benaych-Georges and J.~Rochet, ``\BIBforeignlanguage{English}{Outliers in
  the single ring theorem},'' \emph{\BIBforeignlanguage{English}{Probability
  Theory and Related Fields}}, pp. 1--51, 2015. [Online]. Available:
  \url{http://dx.doi.org/10.1007/s00440-015-0632-x}
\BIBentrySTDinterwordspacing

\bibitem{ipsen2014weak}
J.~R. Ipsen and M.~Kieburg, ``Weak commutation relations and eigenvalue
  statistics for products of rectangular random matrices,'' \emph{Physical
  Review E}, vol.~89, no.~3, p. 032106, 2014.

\bibitem{zhou2013review}
X.~Zhou, S.~Chen, and Z.~Lu, ``Review and prospect for power system development
  and related technologies: a concept of three-generation power systems,''
  \emph{Proceedings of the CSEE}, vol.~22, pp. 1--11, 2013.

\bibitem{mei2014theevolution}
S.~Mei, Y.~Gong, and F.~LIU, ``The evolution model of three generation power
  systems and characteristic analysis,'' \emph{Proceedings of the CSEE},
  vol.~7, pp. 1003--1012, 2014.

\bibitem{he2014power}
X.~He, Q.~Ai, Z.~Yu, Y.~Xu, and J.~Zhang, ``Power system evolution and
  aggregation theory under the view of power ecosystem,'' \emph{Power System
  Protection and Control}, vol.~22, pp. 100--107, 2014.

\bibitem{qiu2011papercog}
R.~C. Qiu, Z.~Hu, Z.~Chen, N.~Guo, R.~Ranganathan, S.~Hou, and G.~Zheng,
  ``Cognitive radio network for the smart grid: experimental system
  architecture, control algorithms, security, and microgrid testbed,''
  \emph{Smart Grid, IEEE Transactions on}, vol.~2, no.~4, pp. 724--740, 2011.

\bibitem{qiu2012efficient}
H.~Li, S.~Gong, L.~Lai, Z.~Han, R.~C. Qiu, and D.~Yang, ``Efficient and secure
  wireless communications for advanced metering infrastructure in smart
  grids,'' \emph{Smart Grid, IEEE Transactions on}, vol.~3, no.~3, pp.
  1540--1551, 2012.

\bibitem{qiu2012scheduling}
H.~Li, L.~Lai, and R.~C. Qiu, ``Scheduling of wireless metering for power
  market pricing in smart grid,'' \emph{Smart Grid, IEEE Transactions on},
  vol.~3, no.~4, pp. 1611--1620, 2012.

\bibitem{IEC2004iec}
D.~Baigent, M.~Adamiak, and R.~Mackiewicz, ``Iec 61850 communication networks
  and systems in substations: an overview for users,'' \emph{The Protection \&
  Control Journal}, vol.~8, pp. 61--68, 2009.

\bibitem{he2013research}
R.~Yuan, Q.~Ai, and X.~He, ``Research on dynamic load modelling based on power
  quality monitoring system,'' \emph{Generation, Transmission \& Distribution,
  IET}, vol.~7, no.~1, pp. 46--51, 2013.

\bibitem{he2014impact}
Q.~Ai, X.~Wang, and X.~He, ``The impact of large-scale distributed generation
  on power grid and microgrids,'' \emph{Renewable Energy}, vol.~62, pp.
  417--423, 2014.

\bibitem{zhang2014economic}
Z.~Hong, D.~Zhao, C.~Gu, F.~Li, and B.~Wang, ``Economic optimization of smart
  distribution networks considering real-time pricing,'' \emph{Journal of
  Modern Power Systems and Clean Energy}, vol.~2, no.~4, pp. 350--356, 2014.

\bibitem{ai2011multi-agent}
Y.~Ji, ``Multi-agent system based control of virtual power plant and its
  application in smart grid,'' Master's thesis, Shanghai Jiaotong University,
  2011.

\bibitem{he2012research}
X.~He, Q.~Ai, P.~Yuan, and X.~Wang, ``The research on coordinated operation and
  cluster management for multi-microgrids,'' in \emph{Sustainable Power
  Generation and Supply (SUPERGEN 2012), International Conference on}.\hskip
  1em plus 0.5em minus 0.4em\relax IET, 2012, pp. 1--3.

\bibitem{ni2007new}
X.~Ni, Q.~Ruan, S.~Mei, and G.~He, ``A new network partitioning algorithm based
  on complex network theory and its application in shanghai power grid,''
  \emph{Power system technology}, vol.~9, pp. 6--12, 2007.

\bibitem{MATPOWER2011matpower}
R.~Zimmerman, C.~Murillo-S{\'a}nchez, and D.~Gan, ``Matpower user¡¯s manual,
  version 4.1,'' \emph{Power Systems Engineering Research Center}, 2011.

\bibitem{zhao2014research}
Y.~Zhao, Y.~An, and Q.~Ai, ``Research on size and location of distributed
  generation with vulnerable node identification in the active distribution
  network,'' \emph{IET Generation, Transmission \& Distribution}, vol.~8,
  no.~11, pp. 1801--1809, 2014.

\bibitem{huang2014impedance}
W.~Huang, T.~Nengling, X.~Zheng, C.~Fan, X.~Yang, and B.~J. Kirby, ``An
  impedance protection scheme for feeders of active distribution networks,''
  \emph{Power Delivery, IEEE Transactions on}, vol.~29, no.~4, pp. 1591--1602,
  2014.

\bibitem{xu2014sparse}
J.~Xu, G.~Yang, Y.~Yin, H.~Man, and H.~He, ``Sparse-representation-based
  classification with structure-preserving dimension reduction,''
  \emph{Cognitive Computation}, vol.~6, no.~3, pp. 608--621, 2014.

\bibitem{pfander2008identification}
G.~E. Pfander, H.~Rauhut, and J.~Tanner, ``Identification of matrices having a
  sparse representation,'' \emph{Signal Processing, IEEE Transactions on},
  vol.~56, no.~11, pp. 5376--5388, 2008.

\bibitem{jana2014fluctuations}
I.~Jana, K.~Saha, and A.~Soshnikov, ``Fluctuations of linear eigenvalue
  statistics of random band matrices,'' \emph{arXiv preprint arXiv:1412.2445},
  2014.

\bibitem{he20153d}
X.~He, Q.~Ai, J.~Ni, L.~Piao, Y.~Xu, X.~Xu \emph{et~al.}, ``3d power-map for
  smart grids---an integration of high-dimensional analysis and
  visualization,'' \emph{arXiv preprint arXiv:1503.00463}, 2015.

\bibitem{he2015unsup}
\BIBentryALTinterwordspacing
X.~He, Q.~Ai, R.~C. Qiu, W.~Huang, and J.~Long, ``An unsupervised learning
  method for early event detection in smart grid with big data,'' \emph{arXiv
  preprint arXiv:1502.00060}, 2015. [Online]. Available:
  \url{http://arxiv.org/pdf/1502.00060.pdf}
\BIBentrySTDinterwordspacing

\bibitem{he2015fault}
Y.~Cao, L.~Cai, C.~Qiu, J.~Gu, X.~He, Q.~Ai, and Z.~Jin, ``A random matrix
  theoretical approach to early event detection using experimental data,''
  \emph{arXiv preprint arXiv:1503.08445}, 2015.

\end{thebibliography}

%
%
%
%
%
%
%
%
\end{document}